\documentclass[twocolumn,trackchanges]{aastex631}
\newcommand{\oiii}{[\ion{O}{3}]}
\newcommand{\mbh}{$M_{\text{BH}}$}
\newcommand{\sigstar}{$\sigma_{\star}$}
\newcommand{\mbhsigstar}{\mbh--\sigstar}
\newcommand{\siggas}{$\sigma_{\text{gas}}$}

\newcommand{\feii}{\ion{Fe}{2}}
\newcommand{\mgib}{\ion{Mg}{1}}

\usepackage{amsmath}

\begin{document}

\title{Spatially Resolved \oiii~Emission Line Kinematics of Reverberation-Mapped AGNs with the Keck Cosmic Web Imager}

\author[0000-0002-0164-8795]{Raymond~P.~Remigio}
\affiliation{Department of Physics and Astronomy, 4129 Frederick Reines Hall, University of California, Irvine, CA 92697, USA}

\author[0000-0002-1912-0024]{Vivian~U}
\affiliation{IPAC, Caltech, 1200 E. California Blvd., Pasadena, CA 91125, USA}
\affiliation{Department of Physics and Astronomy, 4129 Frederick Reines Hall, University of California, Irvine, CA 92697, USA}

\author[0000-0002-3026-0562]{Aaron~J.~Barth}
\affiliation{Department of Physics and Astronomy, 4129 Frederick Reines Hall, University of California, Irvine, CA 92697, USA}

\author[0000-0001-9428-6238]{Nico~Winkel}
\affiliation{Max-Planck-Institut f\"ur Astronomie, K\"onigstuhl 17, D-69117 Heidelberg, Germany}

\author[0000-0003-2064-0518]{Vardha~N.~Bennert}
\affiliation{Physics Department, California Polytechnic State University, San Luis Obispo, CA 93407, USA}

\author[0000-0002-8460-0390]{Tommaso~Treu}
\affiliation{Department of Physics and Astronomy, University of California, Los Angeles, 430 Portola Plaza, Los Angeles, CA 90095, USA}

\author[0000-0001-6919-1237]{Matthew~A.~Malkan}
\affiliation{Department of Physics and Astronomy, University of California, Los Angeles, 430 Portola Plaza, Los Angeles, CA 90095, USA}

\author[0000-0002-6740-411X]{Sebastian~Contreras}
\affiliation{Physics Department, California Polytechnic State University, San Luis Obispo, CA 93407, USA}
\affiliation{Department of Astronomy, San Diego State University, San Diego, CA 92182, USA}

\author[0000-0002-4645-6578]{Peter~R.~Williams}
\affiliation{Department of Physics and Astronomy, University of California, Los Angeles, 430 Portola Plaza, Los Angeles, CA 90095, USA}

\author[0000-0003-4852-8958]{Jordan~N.~Runco}
\affiliation{Department of Physics and Astronomy, University of California, Los Angeles, 430 Portola Plaza, Los Angeles, CA 90095, USA}

\author[0009-0001-9780-4261]{Liam~Hunt}
\affiliation{Department of Physics and Astronomy, 4129 Frederick Reines Hall, University of California, Irvine, CA 92697, USA}

\correspondingauthor{Raymond~P.~Remigio}
\email{remigior@uci.edu}



\begin{abstract}

We present optical integral-field spectroscopic data of ten nearby ($0.02\leq z \leq 0.05$) Seyfert 1 galaxies taken with the Keck Cosmic Web Imager (KCWI). 
We map the spatially resolved kinematics of the \oiii~gas and stars, and investigate the alignments between their global kinematic position angles (PA).
Large-scale gas motions are primarily dominated by rotation, and are kinematically aligned with the stars ($\Delta\text{PA}\leq 30$ deg). 
However, eight galaxies exhibit non-rotational kinematic signatures (e.g., kinematic twists, possible outflows) in their ionized gas velocity fields near the nucleus. We compare aperture-wide measurements of the gas and stellar velocity dispersions (\siggas~and \sigstar) to test the use of the width of the \oiii~line core as a surrogate for \sigstar. 
Direct comparisons between \siggas~and \sigstar~show that \siggas~tends to underestimate \sigstar, and thus is not a reliable tracer of \sigstar~for our selected galaxies.
We measure the extent of the narrow-line region (NLR) using several definitions, resulting in sizes of $\sim0.1$--$10$ kpc. 
For a given \oiii~luminosity, our NLR sizes derived from the [\ion{O}{3}]/H$\beta$ flux ratio or an \oiii~isophotal radius are an order of magnitude larger than those measured from past imaging data.

\end{abstract}



\section{Introduction} \label{sec:intro}

It has been established for decades that supermassive black holes (SMBHs) are an essential component of massive galaxies and are ubiquitous within this population \citep{magorrian98, ho04}.
Extragalactic studies have highlighted correlations between the black hole mass \mbh~and the properties of the host galaxy, including: stellar velocity dispersion \citep[\sigstar,][]{ferrarese00,woo13,kormendy_ho_13,mcconnel_ma_13, woo15, bennert15},
host galaxy luminosity \citep{gultekin09, bentz_manne_nicholas_18, ding20},
bulge luminosity \citep{bentz09, gultekin09},
host stellar mass \citep{ding20},
and bulge stellar mass \citep{marconi_hunt_03, haring_rix_04}.
The correlations between \mbh~and the host parameters suggest a strong coupling and thus co-evolution between the SMBH and the host galaxy (for a review, see e.g., \citealt{kormendy_ho_13}).
Observational studies over the past two decades have solidified the concept of SMBH--host galaxy co-evolution, with analyses of the stars and ionized gas providing insight into specific aspects of this process.
The existence of the \mbhsigstar~relation suggests a link between the growth of the BH and the surrounding bulge \citep{kormendy_ho_13}, while studies of the interstellar medium have shown that active galactic nuclei (AGN) are responsible for driving
ionized gas outflows via radiation pressure and jets \citep[e.g.,][]{mullaney13, woo16, singha22, kim23}. 
Thus, precise measurements of both the ionized gas and the stars are crucial in probing the nature of SMBH--host galaxy co-evolution. 

Of the SMBH--host galaxy scaling relations, \sigstar~has been shown to correlate the tightest with $M_\text{BH}$ \citep{beifiori12, denicola19, bennert21}. 
Measurements of the stellar kinematics involve the use of a template to fit stellar absorption lines in the observed spectrum, which proves challenging in AGN where the host galaxy is outshone by AGN emission.
In particular, contamination from both emission lines (e.g., broad Fe II, [\ion{Fe}{6}], [\ion{Fe}{7}]) and the AGN continuum in the spectral region around the \mgib~triplet makes direct measurements of the bulge stelar velocity dispersion difficult in active galaxies.

To address the difficulties of extracting the stellar kinematics in AGN hosts, the velocity width of the \oiii~profile \siggas~has been used as a tracer for \sigstar, on the grounds that within the bulge, the narrow-line region (NLR) gas kinematics are largely driven by the bulge potential \citep{nelson_whittle_96, Gebhardt_2000, nelson00}. 
More recent works have further studied the use of \siggas~as a surrogate for \sigstar~using high signal-to-noise (S/N) data from the Keck Low Resolution Imaging Spectrometer (LRIS) and the Sloan Digital Sky Survey (SDSS) \citep[e.g.,][]{greene_ho_05, bennert18, sexton21, le23}.
\cite{greene_ho_05} and \cite{bennert18} found that width measurements using the full \oiii~line profile tend to overestimate \sigstar~due to the presence of extended blue wings from outflowing gas.
When only the narrow core of \oiii~(as derived from a two-Gaussian fit) is considered in line width calculations, then \siggas~%closely 
traces
\sigstar, but with large scatter.
Consequently, both studies suggest that \siggas~may be used as a proxy for \sigstar, but only for examining statistical relationships across large samples of galaxies rather than as a direct substitute for individual objects.
However, there are some limitations to previous studies due to differences between the extraction region and the shape of the bulge, as well as assumptions of the bulge's projected morphology.
SDSS spectra correspond to a circular aperture with radius 1\farcs5, which, depending on host galaxy morphology and distance, may extend past the bulge and potentially include contributions from the disk kinematics. 
Spatially resolved long-slit data allow for the spectrum to be integrated out to a specific radius (e.g., the effective radius of the bulge; \citealt{bennert15}), which still implicitly assumes the bulge has a projected circular symmetry.
Testing the validity of the assumption of radial symmetry for bulge velocity dispersions (which is the standard approach) requires 2D spatially resolved kinematic measurements.

The development of integral-field spectroscopy (IFS) has enabled precisely these 2D spatially resolved measurements, revolutionizing the study of gas and stars in galaxies.
Historically, fully characterizing the stars and gas in a galaxy required two distinct approaches: long-slit spectroscopy provided kinematic measurements \citep[e.g.,][]{nelson_whittle_96, bower98, crenshaw00, harris12} and ionization diagnostics \citep[e.g.,][]{baum92}, while imaging data revealed distributions of gas and dust \citep[e.g.,][]{ford94, bennert02, martel04}.
The simultaneous spectral and spatial information enables the analysis of both kinematics and spatial distribution within a single set of observations.
The spatially resolved kinematics enabled by IFS have proven crucial for studying several aspects of SMBH and host galaxy co-evolution, such as AGN-driven outflows and feedback \citep[e.g.,][]{husemann19, riffel21, molina22}, and gas inflows \citep[e.g.,][]{storchi-bergmann10}. 
Recent IFS surveys focusing on nearby AGN \citep[e.g.,][]{mingozzi19, riffel21, husemann22} have revealed the complex interplay between AGN and their host galaxies, providing insights into the stellar and ionized gas kinematics across a range of spatial scales.

In this paper, we present a study of ten local ($ 0.02 \leq z \leq 0.05$) Seyfert 1 galaxies that were observed as part of the Lick AGN Monitoring Project \citep[LAMP,][]{barth15,u22}.
Nine objects have precisely determined black hole masses from classical reverberation mapping \citep[e.g.,][]{u22}, and eight have black hole masses determined from dynamical modeling of the broad-line region \citep[e.g.,][]{villafana22}, both as part of the 2011 and 2016 LAMP campaigns.
This pilot sample will be expanded in a future paper to include 29 Seyfert 1 galaxies
at $z\leq 0.09$ with precisely determined black hole masses spanning several orders of magnitude \citep[e.g.,][]{winkel25}.
Using spectra taken with the Keck Cosmic Web Imager \citep[KCWI,][]{morrissey18}, we generate maps of the spatially resolved stellar and gas kinematics at kiloparsec scales. 
In addition to mapping the stellar and ionized gas kinematics, we investigate differences between the ionized gas and stellar velocity fields at galaxy-wide scales.
We evaluate different methods of quantifying the kinematics within 2D spatial regions in integral-field data, examining the consistency of measurements between 2D spatially resolved and aperture-summed spectra.
Additionally, we investigate the use of \siggas~as a tracer of the bulge potential by comparing the line width of the core component of \oiii~to the stellar velocity dispersion.

The paper is organized as follows. 
We describe the sample and observations in Section \ref{sec:sample}.  
In Section \ref{sec:methods}, we detail the spectral fitting process and
the methods used for measuring our aperture-based velocity dispersions. 
We present our main results for the stellar and gas kinematics in Section \ref{sec:results}.
In Section \ref{sec:nlr}, we present our measurements of the extent of the 
AGN narrow-line region.
We conclude the paper with a summary in Section \ref{sec:summary}. 
Throughout this paper, we adopt $H_0 = 67.8$ km s$^{-1}$ Mpc$^{-1}$, $\Omega_{\text{m}} = 0.308$, and $\Omega_\Lambda = 0.692$ \citep{planck16}.

\section{Sample and Observations} \label{sec:sample}

Our sample consists of ten nearby ($z\leq0.05$)
Seyfert 1 galaxies with reverberation-mapped black hole masses measured as part of the LAMP campaigns.
We list the properties of each galaxy in Table \ref{tbl:sample}.

The targets were observed with KCWI on the Keck II telescope as part of Programs 2018A\_U171 and 2018B\_U012 (PI: Tommaso Treu). 
Our instrument configuration uses the medium slicer and the blue medium (BM) grating, which features a field of view (FoV) of $16\farcs5 \times 20\farcs4$, and wavelength coverage $4645$ \AA~to $5700$ \AA.
With a central wavelength of $\lambda_{\rm{c}} = 5175$ \AA,
our data cover the H$\beta$+\oiii~complex, the \mgib~triplet, and the \ion{Fe}{1} $\lambda$5270 absorption line. From the FeAr arc line exposures, we measure a constant full-width at half-maximum (FWHM) of $0.950$ \AA ~across the entire spectral range, which translates to an instrumental dispersion of $\sigma_{\text{inst}} \approx 24$ km s$^{-1}$.

\begin{figure*}[t]
    \centering
    \includegraphics[width=\textwidth]{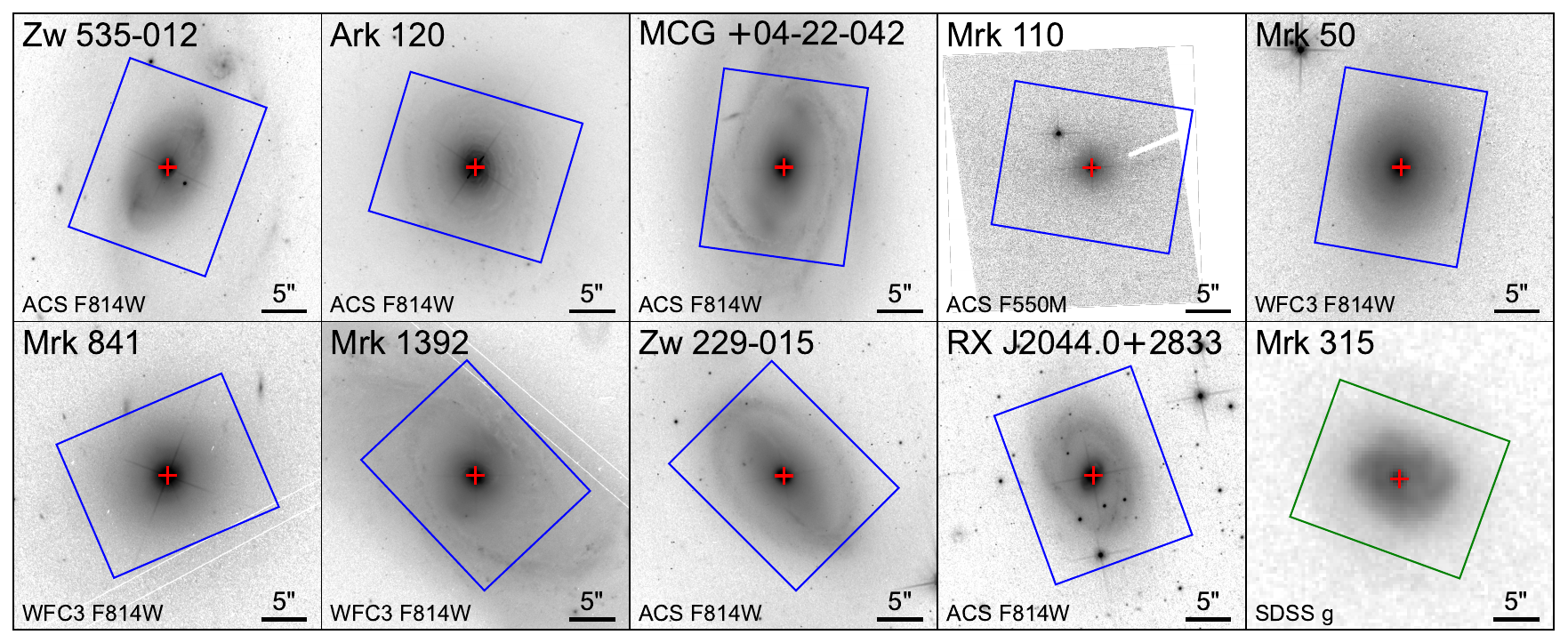}
    \caption{The KCWI FoV plotted as rectangles over optical imaging data from HST (blue) and SDSS (green) for each of the targets. The red plus sign represents the position of the AGN point source, corresponding to the peak of the galaxy emission. North is up and East is to the left.}
    \label{fig:fov}
\end{figure*}

For each target, we chose the observational position angles so that the photometric major axis (as determined from SDSS images) of the host galaxy was aligned with the longer side of the KCWI FoV. The KCWI FoV for each target is displayed in Figure \ref{fig:fov}, superimposed on Hubble Space Telescope (HST) or Sloan Digital Sky Survey (SDSS) images of each galaxy\footnote{Based on observations made with the NASA/ESA Hubble Space Telescope, obtained from the Data Archive at the Space Telescope Science Institute, which is operated by the Association of Universities for Research in Astronomy, Inc., under NASA contract NAS5-26555. These observations are associated with programs 15444, 9851, 16014, and 17103.}.

For each galaxy, we took a short exposure ($\sim 60$ s) to avoid saturation of the bright emission lines (\oiii~and H$\beta$). 
Then, we took a series of longer exposures ($\gtrsim 300$ s), followed by a sky exposure (i.e., an observing pattern of three target exposures, then a sky exposure). 
Typical seeing over our nights was $\sim 1$ arcsec, with one night having a seeing of $\sim 2$ arcsec.

We reduced the data using the Python KCWI Data Reduction Pipeline\footnote{\url{https://github.com/Keck-DataReductionPipelines/KCWI_DRP}}, which includes bias subtraction, flat fielding, wavelength calibration, flux calibration, and datacube creation.
Post-processing tasks were performed through a series of custom Python scripts.
For the datacubes where \oiii~was saturated, we created continuum white light images of both the long and short exposures and used \texttt{image registration}\footnote{\url{https://github.com/keflavich/image_registration}} to calculate pixel offsets. 
After aligning the short exposure datacube with the longer exposure cube, we determined the flux scaling factor from the average flux ratio between cubes near $5100$ \AA. 
The saturated regions in the longer exposure were then replaced with the scaled data from the short exposure. Finally, we used \texttt{image registration} to align the saturation-corrected science frames before co-adding them.
The final datacubes cover a spectral range of $4600$ \AA \textendash $5600$ \AA, with a spectral sampling of $0.5$ \AA$/$px, and rectangular pixels with dimensions $0\farcs679$ $\times$ $0\farcs291$.

\begin{deluxetable*}{lcccccccccc}[htb] 
\tablecaption{Sample Properties and Observations} 
\tablecolumns{11}
\tablewidth{0pt}
\tablehead{ 
\colhead{Object} & 
\colhead{$z$} &
\colhead{$D_L$} &
\colhead{$M_{\text{BH,RM}}$} &
\colhead{$M_{\text{BH,dyn}}$} &
\colhead{$\lambda L_{\lambda}$ (5100 \AA)} &
\colhead{Ref. RM} &
\colhead{Ref. Dyn} &
\colhead{Exp. time} &
\colhead{S/N(AGN)} & 
\colhead{S/N($r_{\text{eff}}$)}
\\
\colhead{} &
\colhead{} &
\colhead{(Mpc)} &
\colhead{($10^8 M_{\odot}$)} &
\colhead{($10^8 M_{\odot}$)} &
\colhead{($10^{43}$ erg s$^{-1}$)} &
\colhead{} &
\colhead{} &
\colhead{(min)} &
\colhead{} & 
\colhead{}
}
\startdata
Zw 535-012 & 0.0478 & 218.2 & 0.38$^{+  0.16}_{-  0.08}$ & \ldots & $4.9 \pm 0.7 $ & 1 & \ldots & 97.5 & 125 & 100 \\ 
Ark 120    & 0.0327  & 148.2 & 0.73$^{+  0.25}_{-  0.17}$ & 1.82$^{+0.58}_{-0.59}$ & $9.2 \pm 3.4$ & 1 & 4 & 80.0 & 150 & 20 \\
MCG $+$04-22-042 & 0.0329  & 150.1 & 0.15$^{+  0.03}_{-  0.02}$ & 0.39$^{+  0.63}_{-0.19}$ & $1.6 \pm 0.4$ & 1 & 4 & 90.0 & 60 & 55 \\
Mrk 110  & 0.0352  & 160.2 & 0.35$^{+  0.06}_{-  0.07}$ & 0.15$^{+0.54}_{-0.07}$ &$7.2 \pm 1.7$ & 1 & 4 & 90.0 & 110 & 75 \\
Mrk 50  & 0.0236  & 107.4 & 0.27$^{+ 0.04 }_{- 0.04}$ & 0.32$^{+0.24}_{-0.11}$ & $0.8 \pm 0.1$ & 2 & 5 & 15.0 & 35 & 5 \\
Mrk 841  & 0.0364  & 165.5 & 0.47$^{+  0.26}_{-  0.16}$ & 0.42$^{+0.90}_{-0.21}$ & $6.7 \pm 0.9$ & 1 & 4 & 90.0 & 150 & 90 \\
Mrk 1392  & 0.0359  & 163.0 & 0.63$^{+  0.08}_{-  0.09}$ & 1.45$^{+0.42}_{-0.38}$ & $1.6 \pm 0.6$ & 1 & 4 & 70.0 & 70 & 55 \\
Zw 229-015 & 0.0278 & 125.9 & 0.09$^{+ 0.02}_{- 0.02}$ & 0.09$^{+0.03}_{-0.03}$ & $0.5 \pm 0.1$ & 3 & 5 & 60.0 & 85 & 35 \\
RX J2044.0$+$2833  & 0.0493  & 229.4 & 0.12$^{+  0.02}_{-  0.02}$ & 0.12$^{+0.05}_{-0.05}$ & $5.3 \pm 0.5$ & 1 & 4 & 75.0 & 85 & 30 \\
Mrk 315$^\text{a,b}$ & 0.0389  & 179.7 & 0.44 & \ldots & $3.2 \pm 0.4$ & \ldots & \ldots & 70.0 & 100 & 35 \\
\enddata \label{tbl:sample}

\tablecomments{
Objects in this and subsequent tables are listed in order of increasing right ascension. Host galaxy redshifts were measured from the Mg$+$Fe absorption line region as a part of this work, and agree to within $ \Delta z = 0.0003$ of the fiducial redshifts as listed in NED.
References for $M_{\text{BH, RM}}$ and $\lambda L_{\lambda}$ (5100 \AA): (1)~\cite{u22}; (2)~\cite{barth11b}; (3)~\cite{barth11a}.
References for $M_{\text{BH,dyn}}$: (4)~\cite{villafana22}; (5)~\cite{williams18}.
Mass measurements from references (2) and (3) were re-scaled to a virial factor of $f=4.47$ to be consistent with (1). 
$^\text{a}$The value of \mbh~for Mrk 315 is an estimate using the single-epoch method of \cite{shen_liu_12}.
The S/N values are the flux density values divided by the 1$\sigma$ uncertainty for a single spectral bin near 5100 \AA, in a single spaxel. S/N(AGN) refers to a spaxel at the nucleus, while S/N($r_{\text{eff}}$) corresponds to a spaxel located at a distance of one effective radius away from the nucleus. $^\text{b}$ We do not have a measurement of the effective radius for Mrk 315, so S/N($r_{\text{eff}}$) is calculated for a spaxel located 1.5 arcseconds from the nucleus.}
\end{deluxetable*}

\section{Spectral Fitting and Kinematic Measurements} 
\label{sec:methods}

\subsection{Host Galaxy Redshift Measurement}

We measured the host galaxy redshift of each target from a fit to an aperture-summed spectrum dominated by starlight.
For each object, we summed the spectra within an annular aperture centered on the AGN point source, where the location of the AGN is determined by the peak flux of a white light image formed by collapsing the data cube across a $5$ \AA~window in the red wing of the broad H$\beta$ line.
We defined the inner radius of the aperture by visually inspecting the spectra in spaxels located far enough from the nucleus such that the H$\beta$ emission from the AGN broad-line region (BLR) was close to zero.
We manually adjusted this radius, incrementally decreasing it to include contributions from higher S/N spaxels.
We chose the inner radius so that the aperture-summed spectra did not show signs of contamination from \feii, [\ion{Fe}{6}], or [\ion{Fe}{7}].
We fitted the aperture-summed spectrum around \mgib~using \texttt{PyPipe3D}\footnote{\url{https://gitlab.com/pipe3d/pyPipe3D/}} \citep{lacerda22} and adopted the best-fit redshift for the remainder of the analyses.
We find that our adopted host galaxy redshifts in the second column of Table \ref{tbl:sample} are in agreement with values listed on the NASA Extragalactic Database (NED)\footnote{The NASA/IPAC Extragalactic Database (NED)
is operated by the Jet Propulsion Laboratory, California Institute of Technology,
under contract with the National Aeronautics and Space Administration.}.

\subsection{Spectral Decomposition}
\label{sub:spectral_fitting}

\begin{figure*}[t]
    \centering
    \includegraphics[width=\textwidth]{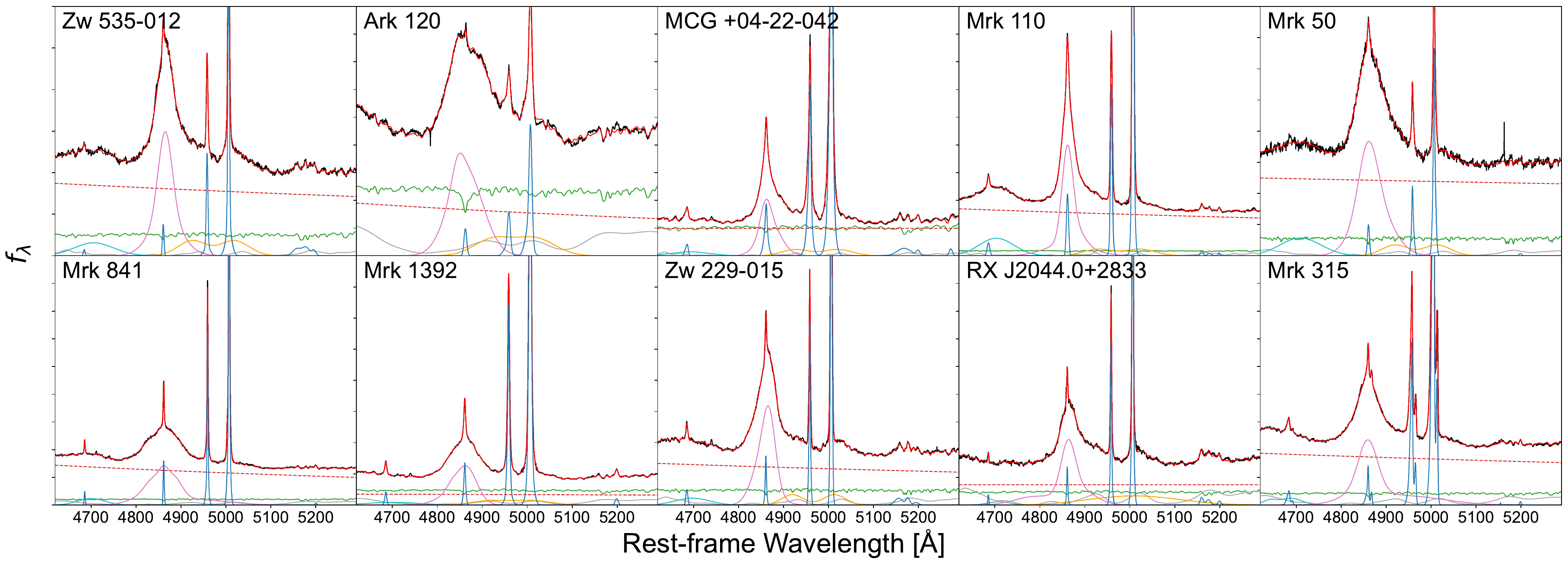}
    \caption{Spectral decompositions illustrating the variety in emission profiles across our sample. The spectral decompositions are for single spaxels near the nucleus, where AGN features (e.g., broad emission lines, power-law continuum, Fe II blends) and the host galaxy could be reliably constrained. Emission associated with the narrow-line region is displayed in blue, host galaxy templates in green, \ion{Fe}{2} templates in gray, and power laws as red dashed lines. Broad emission lines are colored separately, with H$\beta$ in pink, \ion{He}{2} in cyan, and \ion{He}{1} in orange.
    }
    \label{fig:fit}
\end{figure*} 

% the figure set for the spectral decompositions
%\figsetstart
%\figsetnum{3}
%\figsettitle{Spectral Decompositions in Near-Nuclear Spaxels}

%\figsetgrpstart
%\figsetgrpnum{3.1}
%\figsetplot{zw535_decomp.pdf}
%\figsetgrpnote{Spectral decomposition of a near-nuclear spaxel of Zw 535-012. The AGN power law is plotted as a red dashed line, the Fe II template is plotted in gray, and the host galaxy template is plotted in green. Gaussians plotted in pink are associated with broad H$\beta$, cyan with \ion{He}{2}, and orange with \ion{He}{1}. Gaussians associated with narrow lines are plotted in blue (e.g., [\ion{O}{3}], H$\beta$,[\ion{N}{1}]).}
%\figsetgrpend

%\figsetgrpstart
%\figsetgrpnum{3.2}
%\figsetplot{ark120_decomp.pdf}
%\figsetgrpnote{}
%\figsetgrpend

%\figsetgrpstart
%\figsetgrpnum{3.3}
%\figsetplot{mcg_decomp.pdf}
%\figsetgrpnote{}
%\figsetgrpend

%\figsetgrpstart
%\figsetgrpnum{3.4}
%\figsetplot{mrk110_decomp.pdf}
%\figsetgrpnote{}
%\figsetgrpend

%\figsetgrpstart
%\figsetgrpnum{3.5}
%\figsetplot{mrk50_decomp.pdf}
%\figsetgrpnote{}
%\figsetgrpend

%\figsetgrpstart
%\figsetgrpnum{3.6}
%\figsetplot{mrk841_decomp.pdf}
%\figsetgrpnote{}
%\figsetgrpend

%\figsetgrpstart
%\figsetgrpnum{3.7}
%\figsetplot{mrk1392_decomp.pdf}
%\figsetgrpnote{}
%\figsetgrpend

%\figsetgrpstart
%\figsetgrpnum{3.8}
%\figsetplot{zw229_decomp.pdf}
%\figsetgrpnote{}
%\figsetgrpend

%\figsetgrpstart
%\figsetgrpnum{3.9}
%\figsetplot{rxj_decomp.pdf}
%\figsetgrpnote{}
%\figsetgrpend

%\figsetgrpstart
%\figsetgrpnum{3.10}
%\figsetplot{mrk315_decomp.pdf}
%\figsetgrpnote{}
%\figsetgrpend

%\figsetend
% end the figure set

\begin{figure*}
    \centering
    \includegraphics[width=\textwidth]{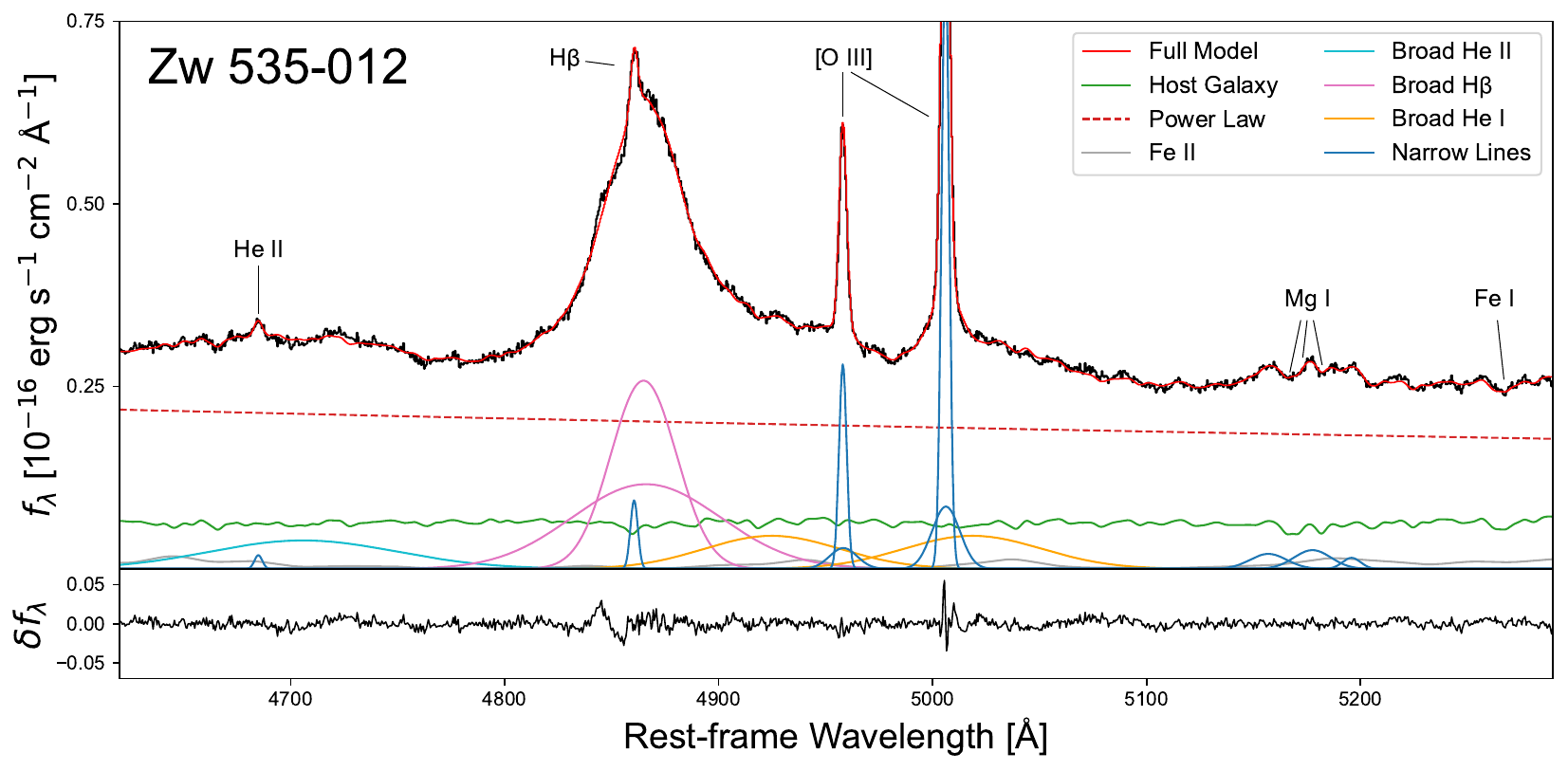}
    \caption{Detailed spectral decomposition of Zw 535-012 for a spaxel near the nucleus. 
    The color coding is similar to Figure \ref{fig:fit}, but with all Gaussians associated with a model component sharing a color (e.g., all Gaussians of broad H$\beta$ are plotted in pink, and all Gaussians associated with narrow lines are plotted in blue). 
    The figure set (10 figures) containing the full spectral decompositions for all targets is available in the online journal.}
    \label{fig:fit2}
\end{figure*}

We used the Bayesian AGN Decomposition Analysis for SDSS Spectra\footnote{\url{https://github.com/remingtonsexton/BADASS3}} (\texttt{BADASS}; \citealt{sexton21}) code to perform our spectral decompositions. 
For each fitting iteration, the \texttt{BADASS} algorithm models and subtracts AGN spectral features (e.g., AGN power-law continuum, \feii~blends, narrow and broad emission lines, absorption lines) from the data spectrum. 
The residual spectrum is assumed to be associated with the host galaxy, which is then fit with a set of stellar template spectra. 
Template spectra (e.g., host galaxy, \feii) were broadened to match the spectral resolution of our data before fitting.
The quality of a fit is assessed from the residuals between the data and the sum of all the model components, as part of a likelihood maximization. 
Observed emission line dispersions are corrected for the instrumental resolution by subtracting off the instrumental dispersion in quadrature from the best-fit line width.

We divided our fitting procedure into three primary steps:
\begin{enumerate}
    \item Create a template spectrum for the spatially unresolved AGN emission.
    \item Fit stellar kinematics.
    \item Fit ionized gas kinematics.    
\end{enumerate}
The AGN template spectrum is used in both subsequent fitting steps, and the stellar kinematics are held constant when trying to fit the \oiii~kinematics.

\subsubsection{AGN Emission Templates} \label{subsec:nuc}

To best characterize both the \oiii~and stellar kinematics, we modeled the spatially unresolved AGN emission, which dominates in nuclear regions.

For each object, we performed a spectral decomposition on an aperture-summed spectrum consisting of spaxels close to the nucleus such that the \feii~and broad H$\beta$ emission are prominent, and spaxels located far enough such that the stellar absorption lines were present. We modeled the AGN continuum using a combination of the \feii~template based on the Seyfert 1 galaxy Mrk 493 \citep{park22} and a power-law continuum with the pivot wavelength fixed to $5100$ \AA. 
To fit the stellar kinematics, we used high resolution (FWHM $=0.550$~\AA) solar metallicity stellar population templates from \cite{maraston_stromback_11}, based on the ELODIE spectral library. 
We note that using these stellar population templates as opposed to the full spectral library may introduce systematic uncertainties of a few percent in measurements of \sigstar~\citep[][]{knabel2025}, though this does not affect the main findings of this work.
We allowed the use of up to
five Gaussians (typically $\sigma > 1000$ km s$^{-1}$) to model the broad H$\beta$ emission, and up to two Gaussians to model the broad \ion{He}{2} $\lambda$4686 emission present in some objects.
Several of our galaxies, most notably Ark 120, display
excess emission in the red wing of broad H$\beta$, referred to in the literature as the ``red shelf''   \citep{meyers_peterson_85, veron02}.
The red shelf contains emission from \feii~and H$\beta$, with possible contributions from
the wings of \oiii~and broad \ion{He}{1}.
Due to the nature of the red shelf, the spectral decompositions for galaxies with this feature are uncertain around \oiii, with several emission features being degenerate with each other.
We found significant residuals in the initial fits to  
most of the galaxies' nuclear spectra in this spectral region, 
so we included contributions from the \ion{He}{1} 4922 \AA ~and 5016 \AA ~lines, which are set to have the same amplitude, velocity offset, and line width \citep{veron02, vestergaard_peterson_05, barth15}.
For two targets (Mrk 841 and Mrk 315), our fits to the nuclear spectrum sufficiently converged without the inclusion of broad \ion{He}{1} emission. However, the resulting fits instead include a highly redshifted component of broad H$\beta$ that covers a similar spectral region.

For each object, we created a template for the AGN spectrum from the best-fit models of the continuum and BLR emission. 
We modified \texttt{BADASS} to use this template as a model component in the subsequent fitting steps, allowing only the amplitude to vary.
Spectral decompositions of near-nuclear spectra separated by components are shown in Figure \ref{fig:fit}.
An example of a full spectral decomposition is shown in Figure \ref{fig:fit2}.

\subsubsection{Stellar Kinematics} \label{subsec:stellar} 

Prior to fitting the stellar kinematics, for each target, we binned the spaxels based on the average signal-to-noise (S/N) ratio of a host galaxy model to the $1\sigma$ error spectrum over a $\sim 20$ \AA ~window centered at $5100$ \AA. 
Subsequent fitting consisted of the following steps.
We first masked out any spaxels where the average  S/N of the data spectrum near $5100$ \AA ~was $<3$. 
Then, we performed a first-pass fit to the stellar kinematics in the unmasked spaxels.
In this fitting run, our model consisted of a host galaxy template, narrow emission lines, and an AGN template created in the preceding section.
On a per-object basis, we incorporated emission lines of [\ion{N}{1}] $\lambda$5200, [\ion{Fe}{6}] $\lambda\lambda$5145,5176, and [\ion{Fe}{7}] $\lambda\lambda$5159,5276, which are prominent in some Type 1 AGN. 
From this fitting run, we generated a new S/N map for each target, where now the S/N is defined as the average flux ratio between the best-fit host galaxy spectrum and the $1\sigma$ error spectrum calculated near $5100$ \AA.
We used these maps in conjunction with the \texttt{VorBin} package \citep{cappellari_copin_03}, choosing a preliminary S/N threshold of 20, adjusting the value on a per-object basis. 
For some objects, (e.g., Mrk 110, Mrk 841, RX J2044.0$+$2833), even the S/N threshold of 20 yielded bins around the nucleus that were much larger than the bulge effective radius.
This is primarily caused by a low S/N of the host galaxy at the nucleus, resulting from the noise contributed by the bright AGN core.
In order to compare the values of \sigstar~obtained from flux-weighted and aperture-summed spectra, we lowered the binning threshold for some objects such that there would be at least two bins where most of the bin area was located within the aperture.
We fit over a spectral region of $4850$~\r{A}~to $5370$~\r{A}~to include the red wing of the broad H$\beta$ emission, which is used to constrain the amplitude of the AGN template.

For several of the nuclear bins in six objects (Zw 535-012, Ark 120, Mrk 110, Mrk 841, RX J2044.0$+$2833, Mrk 315), \texttt{BADASS} failed to converge on a sufficient fit to the host galaxy. 
The fitting software yielded host galaxy models with spurious ($>500$ km s$^{-1}$) velocity dispersions, despite the \mgib~features being visually discernible in the data spectra.
For these bins, we assumed that the continuum level of the host galaxy spectrum can be approximated by a linear function across our wavelength range.
We iteratively scaled the AGN template spectrum, subtracting it from data spectrum until the continuum level of the residual spectrum was linear or flat. 
We then performed a spectral decomposition on the residual spectrum, restricting the fit from 5050 \AA \textendash 5370 \AA ~in order to focus on the \mgib~and \ion{Fe}{1} absorption lines.
In many cases, pre-subtracting the scaled AGN template spectrum yielded measurements of \sigstar~that were more consistent with values in adjacent bins, although these measurements still had large statistical uncertainties.

\subsubsection{\oiii~Kinematics}

\begin{figure*}[htb!]
    \centering
    \includegraphics[width=\textwidth]{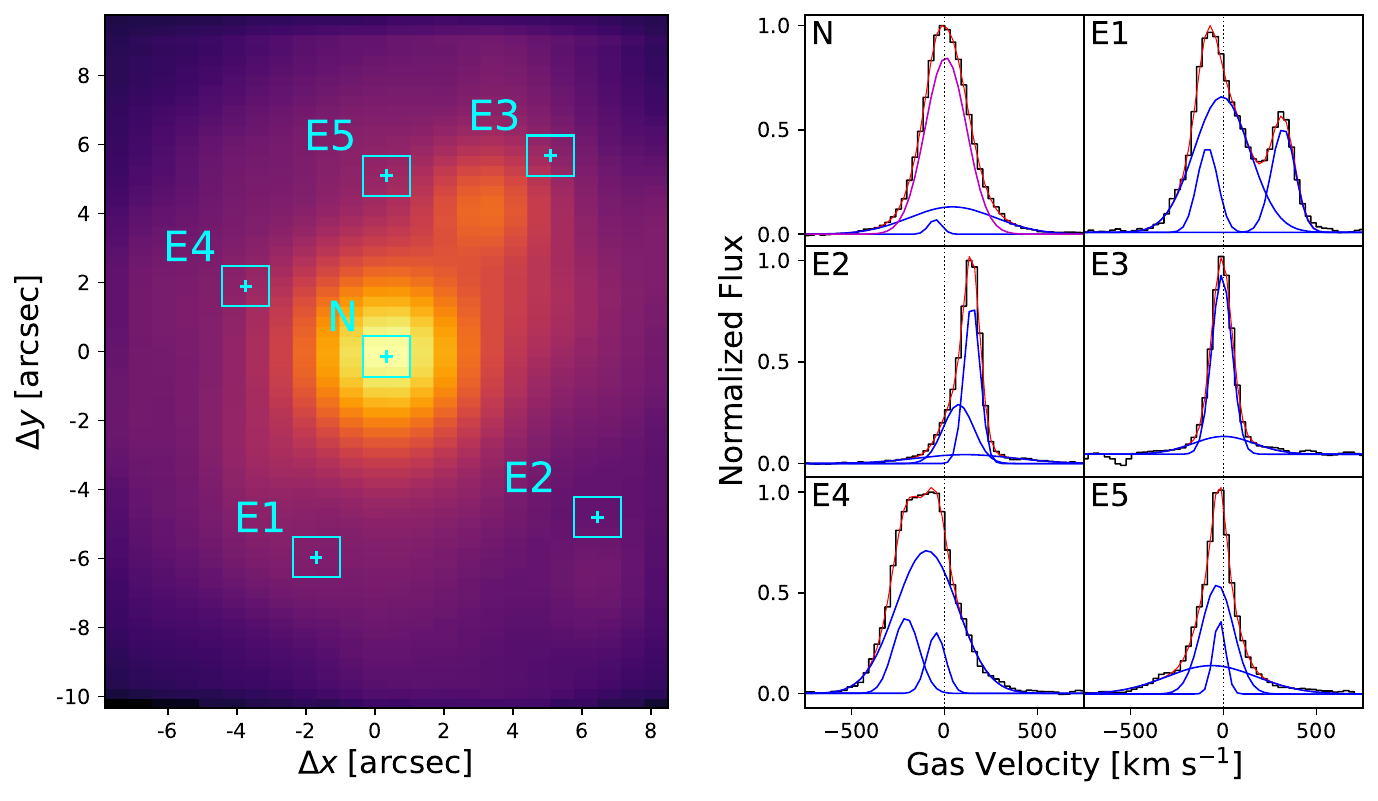}
    \caption{The \oiii~$\lambda$5007 profiles across
    different spatial regions in Mrk 110.
    We chose these regions to display the diversity in the \oiii~line profiles across the FoV.
    The left panel is a white light image generated by collapsing the datacube over a spectral region centered on the \oiii~$\lambda$5007 line. 
    The horizontal axis has been stretched to account for the rectangular pixel sizes in the reduced datacube.
    The $y$-axis of the
    FoV corresponds to a position angle of $80$ deg, which was the orientation of KCWI when we observed this object. The second object near region E3 is a foreground star. 
    In the right panel, we have marked the core component of \oiii~(magenta), which is used in calculations of the bulge velocity dispersion \siggas. Velocities displayed are measured relative to the host galaxy redshift.
    }
    \label{fig:mrk110_profiles}
\end{figure*} 

To extract the ionized gas kinematics, we employed a procedure similar to that of the stellar kinematics.
We perform a first-pass fit to the spectral region around \oiii~$\lambda$5007, where the model only consists of a local, low-order polynomial continuum, and up to four Gaussians for the \oiii~$\lambda 5007$ emission line, depending on the complexity of the profile for a given galaxy.
From this first pass, we generated a map of the S/N of the integrated \oiii~$\lambda$5007 line flux, and masked out any spaxels with S/N$<3$. We used \texttt{Vorbin} to spatially bin the remaining spaxels.

For each spatially binned spectrum, we fixed the stellar velocity and dispersion to values corresponding to the best-fit stellar kinematics of the stellar bin located closest to the centroid of the \oiii~bin.
We performed several passes of fitting over the binned spectra, starting with one Gaussian for the \oiii~profile, with additional iterations with up to four Gaussians. 
We set the velocity widths and offsets for each component of the doublet to be equal to each other, and kept the amplitude ratio of the 5007 \AA\space line to the 4959 \AA\space line fixed at a value of 2.98 \citep{storey_zeippen_00, osterbrock_ferland_2006, dimitrijevic07}. 

To select the best \oiii~line profile fits, we adopted three criteria based on the amplitude of the profile peak, the integrated S/N of the line profile out to $3\sigma$, and the reduced chi-squared ($\chi^2_{\text{r}}$) calculated over a $6 \sigma$ window around the profile centroid.
We adopted a fit with a higher number of Gaussians if the following conditions were satisfied:
\begin{enumerate}
    \item The amplitudes of all \oiii~components were at least three times the flux of the $1\sigma$ error spectrum. 
    \item The integrated S/N of each \oiii~$\lambda$5007 line component is at least 3.
    \item $\chi^2_{\text{r}}$ of the fit with the higher number of Gaussians is less than 0.90$\chi^2_{\text{r}}$ of the fit with the lower number of Gaussians.
\end{enumerate}
In our sample, five objects (Zw 535-012, Ark 120, Mrk 50, Zw 229-015, RX J2044.0$+$2833) required a maximum of two Gaussians to sufficiently describe the \oiii~profile (i.e., a narrow core and a broad wing). 
Mrk 110 and Mrk 315 display complex multi-peaked \oiii~profiles in some spatial regions (see Figure \ref{fig:mrk110_profiles} for Mrk 110), which required three or four Gaussians.
The remaining three objects (MCG$+$04-22-042, Mrk 1392, Mrk 841) required a maximum of three Gaussians to fit \oiii~in some spatial regions, but unlike Mrk 110 and Mrk 315, the third Gaussian was often needed to accurately fit the wings of the \oiii~profile.
The addition of the third Gaussian in these objects makes it difficult to interpret whether the additional Gaussian is a distinct kinematic component, or if its simply needed to describe a non-Gaussian core or wing component.
Given these interpretational challenges, for all objects (except Mrk 110 and Mrk 315), we adopted the two-Gaussian fits in spatial bins where three Gaussians were determined to be statistically significant. 
This approach aligns with previous literature investigating the use of \oiii~as a proxy for \sigstar~\citep{greene_ho_05,bennert18}, where the narrow line core is typically associated with gravitational motion, and a broader wing component is indicative of non-rotational motion (e.g., turbulence, inflows, outflows).
For Mrk 110, most spaxels within the bulge required three Gaussians to fit the \oiii~profile, with two narrow components and a single broad base. 
For the purpose of calculating \siggas~for Mrk 110, we interpreted the higher amplitude narrow Gaussian as the rotational component.

\subsection{Moment Maps and Kinematic PA Measurements}

For each object, we produced maps of the \oiii~surface brightness, velocity, and dispersion, and the stellar kinematics.
Since the \oiii~profile is non-Gaussian for
most objects in the sample, the integrated line flux, velocity, and dispersion are obtained from the zeroth, first, and second moments, respectively:
\begin{equation} \label{eq:F}
    F = \int_{\text{line}}f_\lambda(\lambda) d\lambda
\end{equation}
\begin{equation} \label{eq:lamc}
    \lambda_{\text{c}} = \frac{\int_{\text{line}}\lambda f_\lambda(\lambda) d\lambda}{F}
\end{equation}
\begin{equation} \label{eq:sig2}
    \sigma^{2}_{\lambda} = \frac{\int_{\text{line}} (\lambda-\lambda_c)^2 f_\lambda(\lambda) d\lambda}{F},
\end{equation}
where the first and second moments are converted into velocities by
\begin{equation} \label{eq:vc}
    v_c = \left(\frac{\lambda_c}{\lambda_0} - 1\right) c
\end{equation}
\begin{equation}
    \sigma_{v} = \frac{\sigma_{\lambda}}{\lambda_c} c,
\end{equation}
where $\lambda_0$ is the rest wavelength of the [\ion{O}{3}] line.
For some objects where a non-rotational component may be isolated from the \oiii~profile (Ark 120, MCG$+$04-22-042, Mrk 841, Mrk 1392), we generated maps of the surface brightness, velocity, and dispersion of these components. 
In the cases where this non-rotational component could not be easily separated across the entire galaxy (e.g., Mrk 315 and Mrk 110), we produced maps of the \oiii~surface brightness within 60 km s$^{-1}$ velocity slices.
We use the \texttt{astropy} package \texttt{reproject} to resample the maps to a standard WCS system with $0\farcs679$ square pixels, corresponding to the larger dimension of the pixels in the reduced datacubes.

We measured and compared the global kinematic position angles from the resampled \oiii~and stellar velocity maps using \texttt{PaFit}\footnote{\url{https://www-astro.physics.ox.ac.uk/~cappellari/software/}} \citep{krajnovic06} (see Appendix C of that paper for a detailed explanation of the algorithm). 
Briefly, the best-fit global kinematic position angle (PA) is selected by minimizing the residuals between the galaxy's generated velocity field (which resembles disk rotation) and the maps generated from our analysis.
For all objects aside from Mrk 315, we measured the kinematic PA based on the velocity maps as generated using Equations (\ref{eq:lamc}) and (\ref{eq:vc}).
For Mrk 315, where there is extended emission from highly redshifted \oiii~($v\gtrsim 500$ km s$^{-1}$), we provided a velocity map created from spaxels dominated by rotation.
We provided \texttt{PaFit} with a coarse estimate of the uncertainties in measured velocity, based on the statistical uncertainties output by \texttt{BADASS}. 
We took the median of the velocity uncertainties over all the fitted bins for a given target.
For spaxels where multiple Gaussians were required to fit \oiii, we defined the velocity uncertainty in that bin to be the quadrature sum of the velocity uncertainties for each component.

The kinematic maps and PA measurements will be discussed in Section \ref{sec:results}.

\subsection{Flux-weighted and Aperture-summed $\sigma$ Measurements} \label{subsec:sig_measure}

To investigate the viability of \siggas~as a surrogate for the stellar dispersion \sigstar, we performed additional measurements on the stellar and gas dispersion within an elliptical aperture centered on the AGN point source. 
We used the \texttt{lenstronomy} \footnote{\url{https://github.com/lenstronomy/lenstronomy}}
\citep{birrer_amara_18, Birrer2021} package to model the surface brightness profile of all galaxies with available HST images and obtain measurements of effective radius $r_{\text{eff}}$, axis ratio $q$, and the position angle $\theta$ of the S\'ersic spheroid denoting the bulge. 
We employed the same procedure outlined in \cite{bennert21}, and will present detailed results on the surface brightness photometry (for objects with high resolution HST imaging data available) in a companion paper (Bennert et al.\:2025, in prep).
The fitted parameters are used to define an elliptical aperture within which we measure \siggas~and \sigstar, with $r_{\text{eff}}$ taken to be the semi-major axis. 

We measured the velocity dispersion within the elliptical aperture through two different methods.
In the first method, we calculated a flux-weighted dispersion $\sigma_{\text{w}}$ based on the spectral fits to the individual bins located within the elliptical aperture:
\begin{equation} \label{eq:sig}
    \sigma^{2}_{\text{w}} = \frac{\sum\limits_{i=1}\limits^{N} (\sigma^{2}_{i} + v^{2}_{i}) F_{i} \alpha_i}{\sum\limits_{i=1}\limits^{N} F_{i} \alpha_i},
\end{equation}
where $F_{i}$ is the flux within the $i$-th bin, and $\alpha_i$ is an additional weighting factor between $0$ and $1$ that corresponds to the fraction of the $i$-th bin that lies within the aperture. 
This definition is similar to measurements of $\sigma$ derived from spatially resolved long-slit spectroscopy \citep[e.g.,][]{gultekin09, mcconnel_ma_13, kormendy_ho_13, bennert15}, but we directly weight $(\sigma^2 + v^{2})$ by the flux in each spaxel,
rather than weighting by a radial intensity profile and assuming circular symmetry. 
For the \oiii~profile, $v_i$ and $\sigma_i$ are the line velocity and dispersion of the core component.
For most of our sample, we define the line core as the narrower component in a two-Gaussian fit to the \oiii~ profile.
For Mrk 110, which requires three Gaussians to describe the \oiii~profile, we interpret the highest amplitude component as the line core (see Figure \ref{fig:mrk110_profiles}).
In the second method, we summed up the spectra in all spaxels located within the elliptical aperture, once again applying the weighting $\alpha_i$ to take into account spaxels located partially inside the aperture.
We then used \texttt{BADASS} to perform a fit to the aperture-summed spectrum and directly obtain values of \siggas~and \sigstar. 
We were unable to measure \sigstar~for two galaxies (Zw 535-012 and RX J2044.0$+$2833), since the sizes of the nuclear Voronoi bins for the stellar kinematics exceed the bulge effective radius.

To investigate potential systematic effects from our choice of aperture, we performed additional measurements using circular apertures of radius $r_{\text{eff}}$ and rectangular apertures designed to simulate long-slit observations.
For the rectangular aperture measurements, we isolated and Voronoi binned the spaxels along a single slice in the datacube passing through the nucleus. 
Since the photometric major axis of each galaxy was oriented along slices, these spaxels simulated an effective extraction window with a width of 0\farcs679.
 
\begin{deluxetable*}{lccc}[hb]
\tablecaption{Global Kinematic PAs}
\tablecolumns{4}
\tablewidth{3pt}
\tablehead{
\colhead{Object} & 
\colhead{$\text{PA}_{\text{gas}}$} &
\colhead{$\text{PA}_{\star}$} &
\colhead{$\Delta \text{PA}$} \\
\colhead{} &
\colhead{(deg)} &
\colhead{(deg)} &
\colhead{(deg)}
}
\startdata
Zw 535-012        & $174.6\pm1.4$ & $185.9\pm2.5$\phn   &  $11.3\pm2.9$\phn\\
Ark 120           & \phn$15.0\pm3.4$  & \phn$11.0\pm1.6$\phn    &  \phn$4.0\pm3.8$\phn\\
MCG $+$04-22-042  & $184.1\pm0.5$ & $182.7\pm0.8$\phn  &  \phn$1.4\pm0.9$\phn\\
Mrk 110           & $305.6\pm2.9$ & $309.7\pm25.1$  &  \phn$4.1\pm25.3$\\
Mrk 50            & $334.8\pm8.2$ & $352.1\pm7.4$\phn   &  $17.3\pm11.0$\\
Mrk 841           & $137.3\pm5.1$ & $132.8\pm8.6$\phn   &  \phn$4.5\pm10.0$\\
Mrk 1392          & $235.7\pm2.5$ & $234.4\pm1.5$\phn   &  \phn$1.3\pm2.9$\phn\\
Zw 229-015        & $226.8\pm4.5$ & $223.4\pm3.2$\phn   &  \phn$3.4\pm5.5$\phn\\
RX J2044.0$+$2833 & \phn$13.7\pm0.5$  & \phn\phn$9.0\pm1.0$\phn    &  \phn$4.7\pm1.1$\phn\\
Mrk 315           & \phn$69.0\pm0.5$  & \phn$62.0\pm3.0$\phn    &  \phn$7.0\pm3.0$\phn\\
\enddata \label{tbl:pa}
\tablecomments{Angles listed for PA are measured counter-clockwise (East) from North to the redshifted section of the velocity field. The uncertainties for the PA$_{\text{gas}}$ and PA$_{\star}$ measurements are those returned directly from \texttt{PaFit}. These values were added in quadrature to obtain the uncertainties for $\Delta$PA.} 
\end{deluxetable*}

\movetabledown=2in
\begin{rotatetable*}
\begin{deluxetable*}{lccccccccccccc}
\tablecaption{\texttt{lenstronomy} Results and Aperture-based $\sigma$ Measurements}
\tablecolumns{14}
\tablewidth{3pt}
\tablehead{
\colhead{Object} & 
\colhead{$r_{\text{eff}}$} &
\colhead{$r_{\text{eff}}$} &
\colhead{$q$} &
\colhead{$\theta$} &
\colhead{$\sigma_{\text{gas,w}}$} &
\colhead{$\sigma_{\star, \text{w}}$} &
\colhead{$\sigma_{\text{gas,a}}$} &
\colhead{$\sigma_{\star, \text{a}}$} &
\colhead{$\sigma^{\text{ls}}_{\text{gas,w}}$} &
\colhead{$\sigma^{\text{ls}}_{\star, \text{w}}$} & 
\colhead{$\sigma^{\text{ls}}_{\text{gas,a}}$} &
\colhead{$\sigma^{\text{ls}}_{\star, \text{a}}$} &
\colhead{FWHM/2.355} \\
\colhead{} &
\colhead{(arcsec)} &
\colhead{(kpc)} &
\colhead{} &
\colhead{(deg)} &
\colhead{(km s$^{-1}$)} &
\colhead{(km s$^{-1}$)} &
\colhead{(km s$^{-1}$)} &
\colhead{(km s$^{-1}$)} &
\colhead{(km s$^{-1}$)} &
\colhead{(km s$^{-1}$)} &
\colhead{(km s$^{-1}$)} &
\colhead{(km s$^{-1}$)} &
\colhead{(km s$^{-1}$)}
}
\startdata
Zw 535-012        & 0.58 & 0.56 & 0.80 &172.8 &$106 \pm 2$    & \phn$\cdots$ & $101 \pm 1$  & \phn$\cdots$ & \phn$91\pm4$ & \phn$\cdots$ & $101\pm1$ & \phn$\cdots$ & $110\pm1$ \\
Ark 120           & 1.99 & 1.34 & 0.87 &\phn\phn2.2 &$208 \pm 1$    & $203 \pm 4$\phn & $218 \pm 7$  & $135\pm2$ & $217\pm3$ & $229\pm8$\phn & $189\pm9$ & $259\pm20$  & $217\pm6$  \\
MCG $+$04-22-042  & 0.93 & 0.63 & 0.78 &170.8 &$110 \pm 1$    & $171 \pm 3$\phn & \phn$61 \pm 2$  & $138 \pm 2$ & $110\pm1$ & $180\pm5$\phn & $109\pm2$ & $110\pm2$\phn & $83\pm4$ \\
Mrk 110           & 1.53 & 1.10 & 0.97 &\phn\phn3.8 &$106 \pm 1$    & $104 \pm 5$\phn & \phn$113 \pm 1$  & \phn$86 \pm 3$ & $107\pm1$ & $137\pm29$ & \phn$82\pm1$ & $87\pm7$ & $127\pm1$  \\
Mrk 50            & 4.15 & 2.04 & 0.80 &171.3 &\phn$75 \pm 1$ & \phn$95  \pm 1$\phn & \phn$58  \pm 3$ & \phn$85  \pm 3$ & \phn$81\pm2$ & $90\pm2$ & \phn$54\pm4$ & $92\pm4$ & $102\pm4$ \\
Mrk 841           & 1.30 & 0.97 & 0.95 &104.8 &\phn$54  \pm 1$& $102 \pm 8$\phn & \phn$56  \pm 1$  & $104 \pm 4$ & \phn$55\pm1$ & $80\pm7$ & \phn$54\pm1$ & $90\pm7$ & $66\pm1$ \\
Mrk 1392          & 0.74 & 0.55 & 0.76 &\phn40.7 &$101 \pm 2$    & $154 \pm 9$\phn & $100  \pm 1$  & $151 \pm 4$ & $102\pm2$ & $137\pm5$\phn & $139\pm3$ &  $141\pm3$\phn & $121\pm4$ \\
Zw 229-015        & 0.75 & 0.43 & 0.79 &\phn44.9 &\phn$59  \pm 1$& $85  \pm 3$ & \phn$34  \pm 1$  & \phn$79  \pm 7$ & \phn$60\pm1$ & $88\pm6$ & \phn$33\pm1$ & $57\pm8$ & $73\pm2$ \\
RX J2044.0$+$2833 & 0.24 & 0.24 & 0.89 &\phn10.3 &\phn$69 \pm 1$ & \phn$\cdots$ & \phn$49  \pm 1$  & \phn$\cdots$ & \phn$71\pm2$ & $\cdots$ & \phn$81\pm2$ & \phn$\cdots$ & $68\pm2$ \\
\enddata  \label{tbl:kin}
\tablecomments{
Columns 2 through 5 (effective radius in arcsec and kpc, axis ratio, and orientation) are the best-fit values of the spheroidal S\'ersic component for each galaxy, and are used to define the elliptical aperture where the velocity dispersions are calculated.
Bulge position angles $\theta$ listed in the fourth column are measured counter-clockwise (East) from North. The spheroid measurements are robust to within 10 percent. The values of $\sigma_{\text{w}}$ are calculated using Equation (\ref{eq:sig}), and
values of $\sigma_{\text{a}}$ and and FWHM/2.355 are obtained by a fit to the aperture-summed spectra. Uncertainties for $\sigma_{\text{w}}$ were obtained by propagating the uncertainties for each individual bin, while uncertainties for $\sigma_{\text{a}}$ and FWHM/2.355 are directly from \texttt{BADASS}.}
\end{deluxetable*}
\end{rotatetable*}

\section{Stellar and Gas Kinematics}{\label{sec:results}}

\begin{figure*}[htb!]
    \centering
    \includegraphics[width=\textwidth]{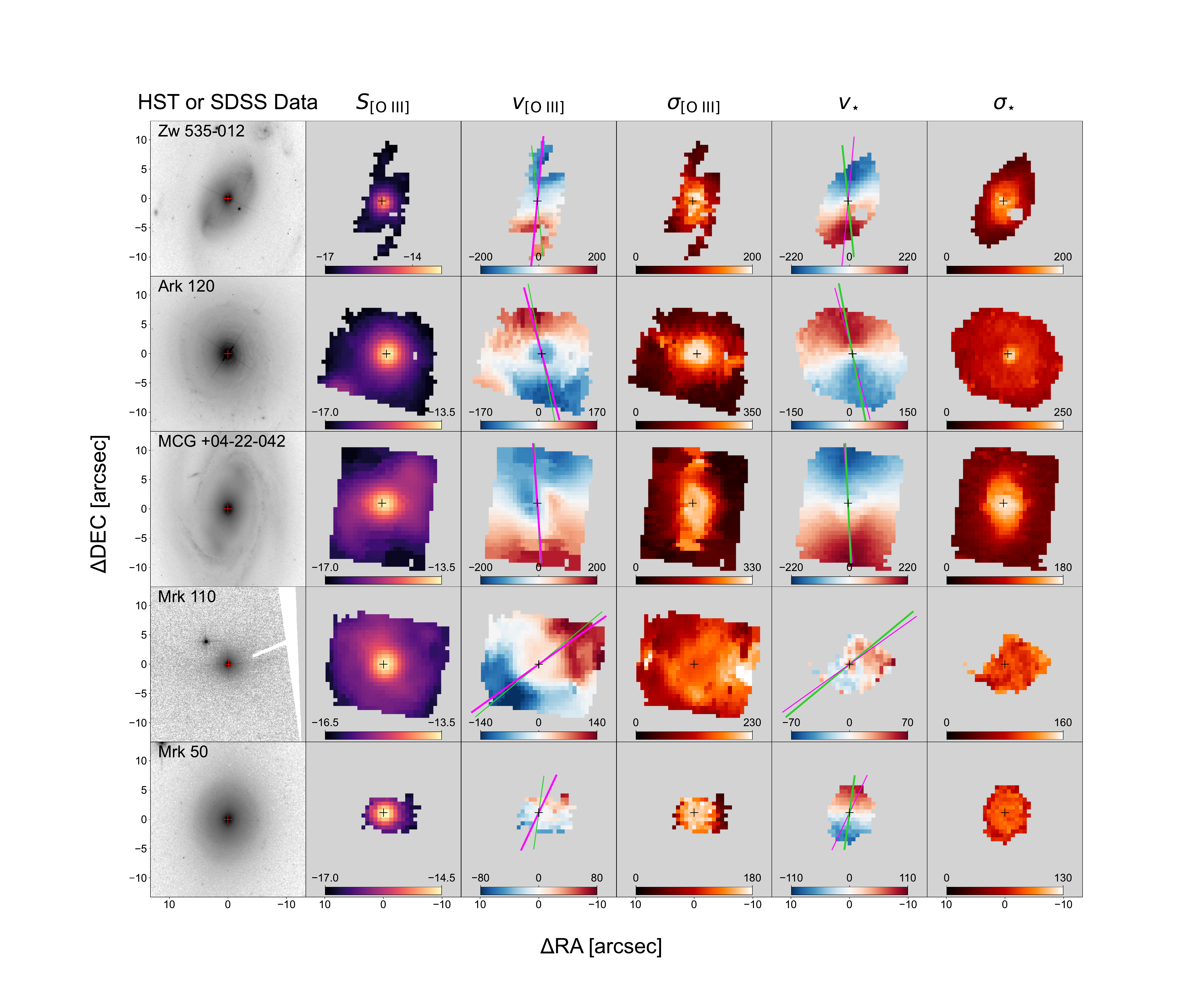}
    \caption{Optical HST (SDSS for Mrk 315) image together with three moment maps (i.e., surface brightness, centroid velocity, and velocity dispersion) of the ionized gas component, and the kinematic maps ($v_{\star}$, $\sigma_{\star}$) of the stellar component, for five objects in the sample as derived from our KCWI data.
    Images used in the first column are the same as those from Figure \ref{fig:fov}, but resampled with North up and East to the left. The red plus sign denotes the location of the AGN point source, as derived from the peak emission in the imaging data. In other panels, the black plus sign denotes the location where the AGN continuum emission is brightest. The surface brightness is plotted in units of $\log_{10} S_{\text{[O III]}}$, while velocity and dispersion are plotted in units of km s$^{-1}$. The magenta and green lines on the velocity maps denote the measured PA of \oiii~and the stars, respectively. In each velocity map, the thicker line corresponds to the PA of the velocity field shown in that panel. We masked out several spaxels in Zw 535-012 and Mrk 110 where the spectra were dominated by foreground stars.}
    \label{fig:full_a}
\end{figure*} 

\begin{figure*}[htb!]
    \centering
    \includegraphics[width=\textwidth]{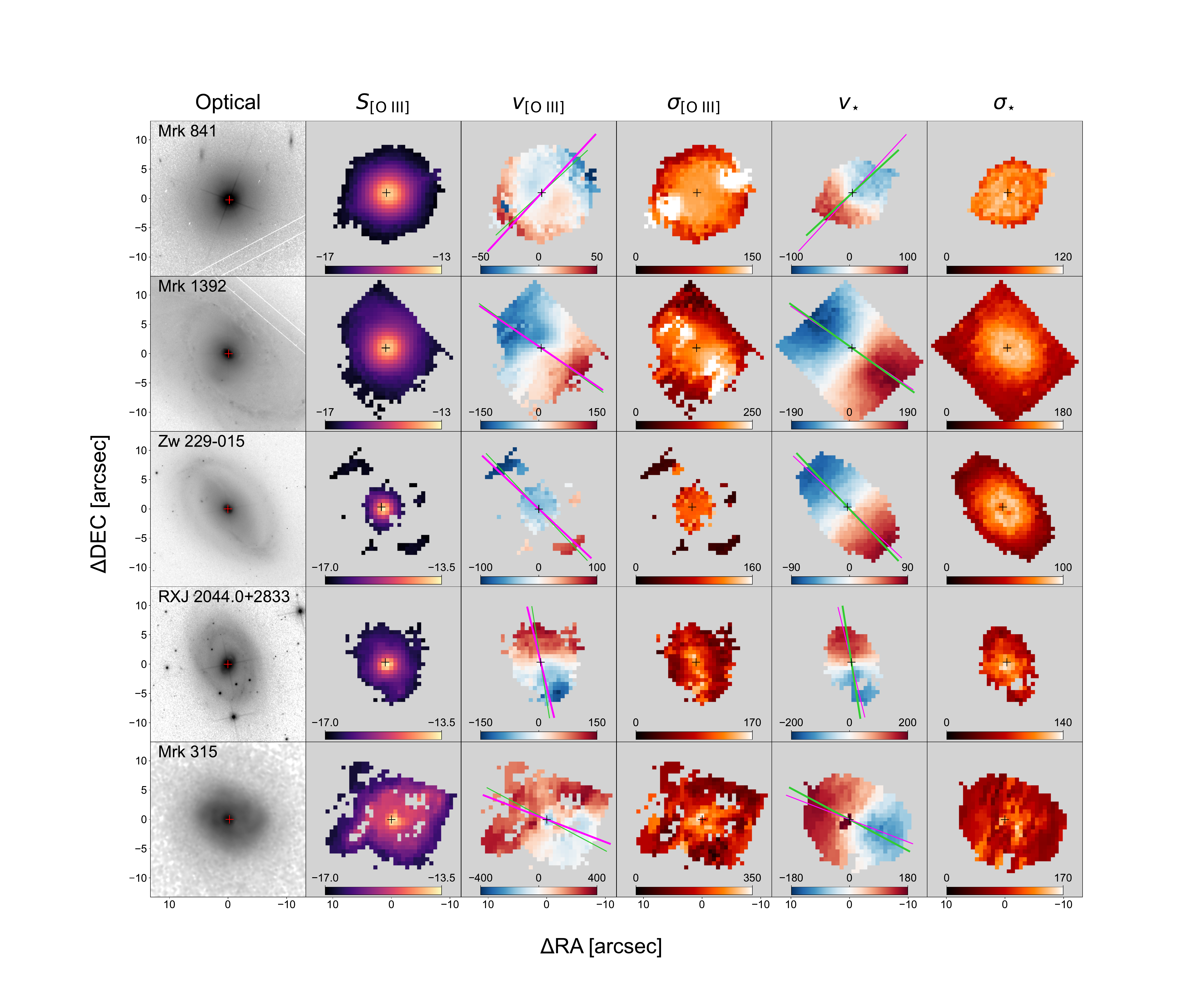}
    \caption{Same as Figure \ref{fig:full_a}, but for the remaining objects in the sample. The masked regions in RX J2044.0$+$2833 correspond to foreground stars located $\lesssim5$ arcsec south of the nucleus.}
    \label{fig:full_b}
\end{figure*} 

\begin{figure*}[htb!]
    \centering
    \includegraphics[width=\textwidth]{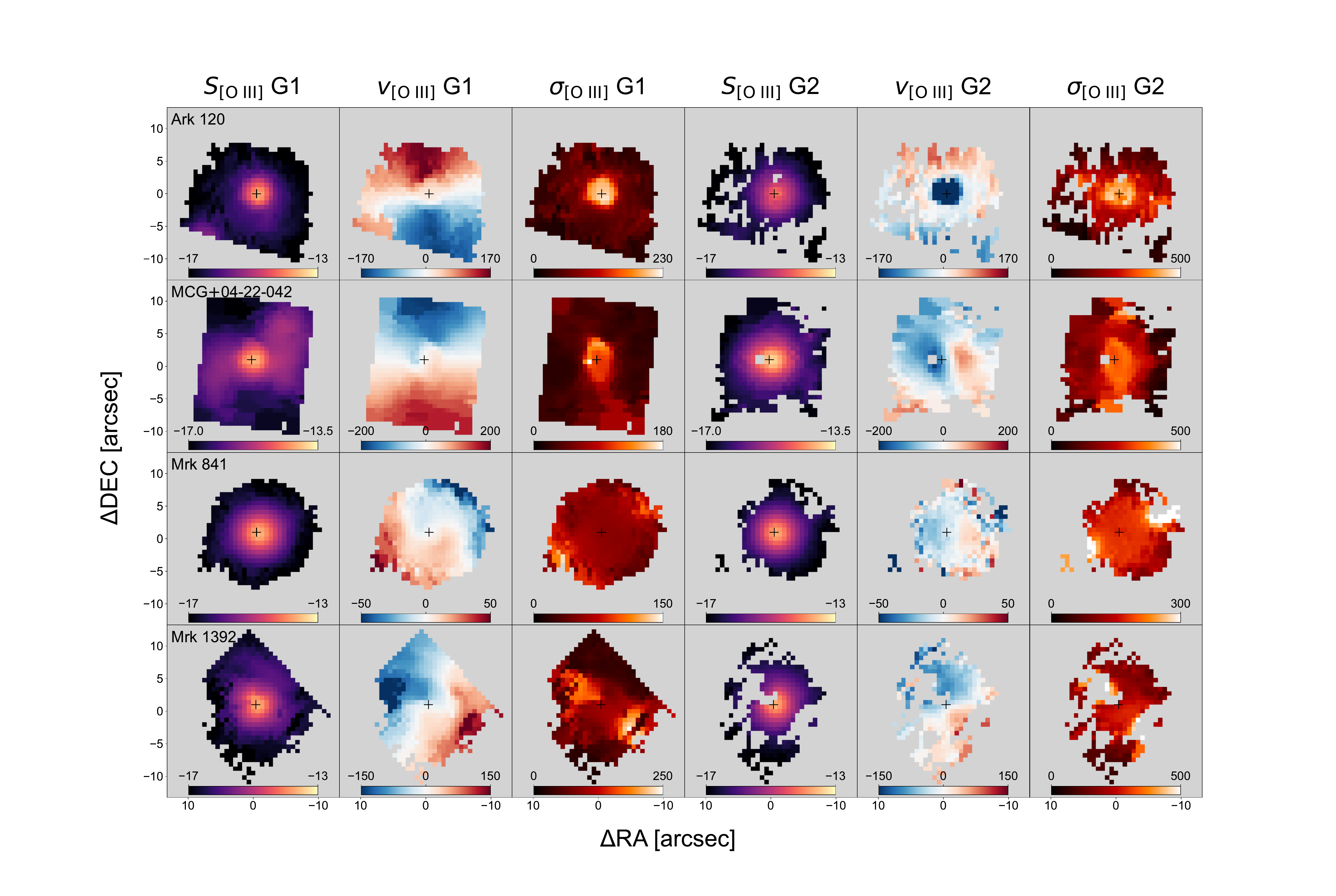}
    \caption{Surface brightness and ionized gas kinematics for the four galaxies with separable components. Component G1 refers to the narrow core of the \oiii~profile, while G2 refers to the broad asymmetric wing.}
    \label{fig:sep_all}
\end{figure*} 

\subsection{Individual objects}
We present the moment maps for the stellar and ionized gas kinematics for our sample in Figures \ref{fig:full_a} and \ref{fig:full_b}.
The velocity maps show the global kinematic PA for both the gas (magenta) and stars (green) as measured from \texttt{PaFit} (Table \ref{tbl:pa}).  
Our aperture-based measurements of $\sigma$ are listed in Table \ref{tbl:kin}.

We find that for all objects, the differences in the kinematic PAs are $\Delta \rm{PA} \lesssim 20$ deg, which is well below the typical $\Delta \text{PA}=30$ deg threshold used to denote gas-star misalignments \citep[e.g.,][]{jin16, bryant19}.
It should be noted that the $\text{PA}_{\text{gas}}$ reported in these results captures the larger-scale rotation of \oiii. 
Thus, we can conclude that at larger (kpc) scales, the stars and the gas are kinematically aligned.
Nevertheless, most objects show non-rotational signatures in their \oiii~kinematics at smaller scales. 
Six targets (Ark 120, MCG$+$04-22-042, Mrk 841, Mrk 1392, Mrk 110, and Mrk 315) exhibited \oiii~profiles that had asymmetric wings or multiple peaks. 
For the four galaxies (Ark 120, MCG$+$04-22-042, Mrk 841, Mrk 1392) where we could sufficiently distinguish between two kinematically distinct
components (e.g., a narrow line core and a broad asymmetric wing), we display the \oiii~moment maps for each component separately in Figure \ref{fig:sep_all}.
We denote the narrow core as component G1, and the broad wing as G2.
For these galaxies, G1 displays velocity dispersions up to $\sim150$\textendash$250$ km s$^{-1}$, while G2 has velocity dispersions up to $\sim500$ km s$^{-1}$.

\textbf{Zw 535-012:} There are two patches of elevated \oiii~velocity dispersion that are aligned with the photometric major axis. The dispersion peaks are located 2.5 arcsec (2.4 kpc) away from the nucleus. 
The distribution of \oiii~extends into two smaller ``arms'' towards the north east and south west.

\textbf{Ark 120:} The \oiii~profile shows a highly blueshifted ($v \sim -500$ km s$^{-1}$) wing that is prominent in spaxels within $\sim 2$ arcsec (1.4 kpc) of the nucleus. 
We suspect that this could be a face-on AGN-driven outflow, but we also consider the possibility that this could be strong \feii~emission based on the galaxy's distinct ``red shelf'' in its spectrum and that the morphology of the blueshifted gas resembles a point source. 
We also note an area of faint, narrow double-peaked emission located 10 arcsec (6.7 kpc) towards the south east of the nucleus.

\textbf{MCG $+$04-22-042:} The \oiii~velocity map shows a distinct kinematic twist near the nucleus, with larger scale motions being nearly aligned with the stellar kinematics. 
At distances within 5 arcsec (3.4 kpc), \oiii~velocity field is misaligned from the stellar velocity field by much as $\Delta \rm{PA} \lesssim 70 \; \rm{deg}$.
We find that the gas dispersion shows an area of enhanced line width that extends along the major axis of the galaxy, tracing the morphology of the bar seen in the imaging data.  
The narrower core component (G1) displays a slight kinematic twist near the nucleus, but otherwise shows disk rotation.  
The highest velocities of the non-rotational component (G2) are located near the nucleus, and are oriented roughly East-West.
The feature is oriented nearly perpendicular to the global kinematic PA ($\Delta \; \rm{PA} = 90.3 \pm 0.5 \; \rm{deg}$), as well as the bar.
One possible interpretation of the second component is an East-West AGN driven outflow, while at larger spatial scales, the second component is tracing out the spiral arms.
Similar to the dispersion map generated from the full profile, we find that the dispersion map of the non-rotational component traces the bar.

\begin{figure*}[htb!]
    \centering
    \includegraphics[width=\textwidth]{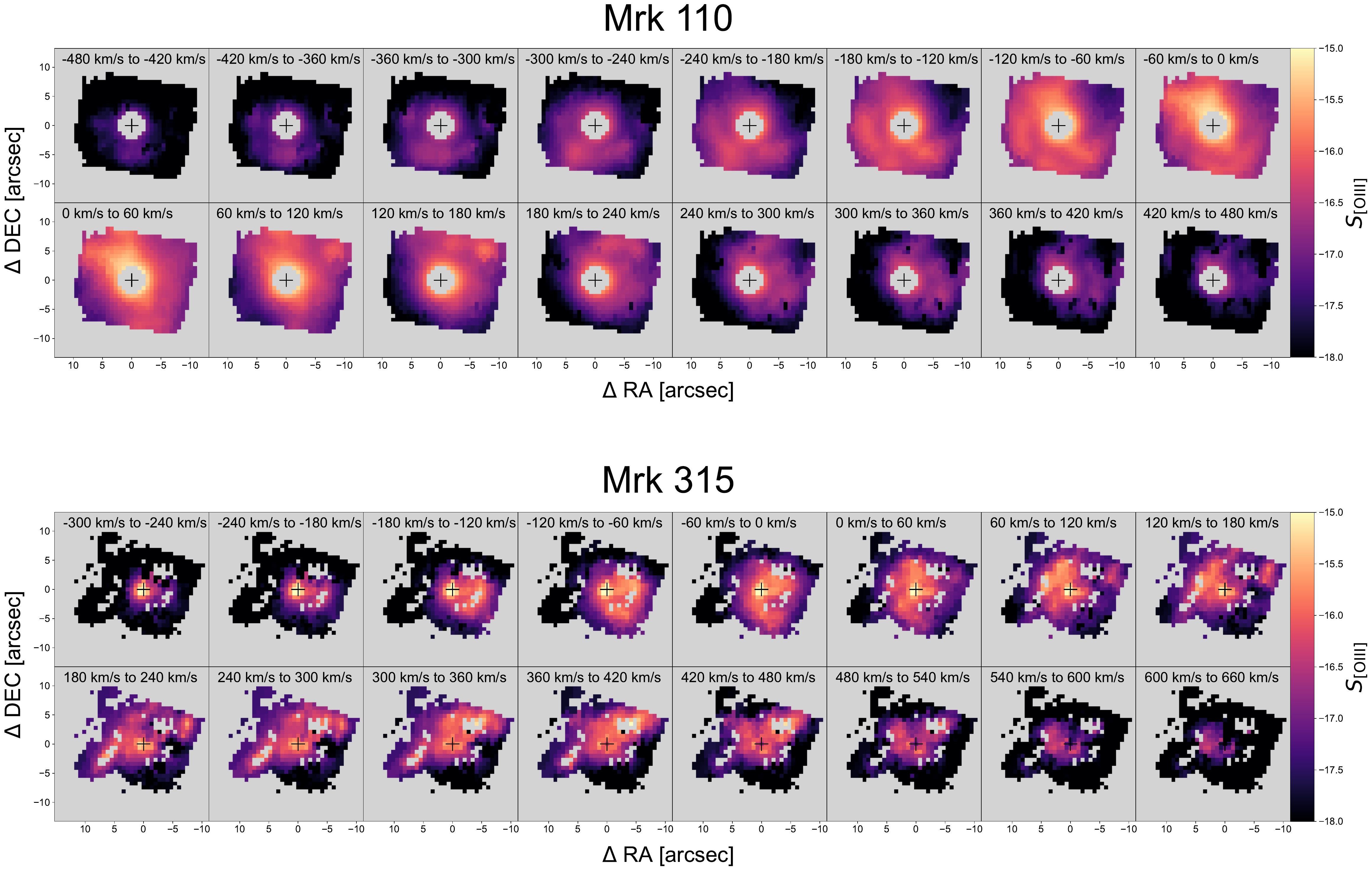}
    \caption{\oiii~Surface brightness maps of Mrk 110 and Mrk 315 within 60 km s$^{-1}$ bins. The inner 2 arcsec of Mrk 110 are masked to better display the extended \oiii~emission. Similar to Figures \ref{fig:full_a} and \ref{fig:full_b}, surface brightness is plotted in units of $\log_{10} S_{\mathrm{[O\;III]}}$.}
    \label{fig:velslice}
\end{figure*} 

\textbf{Mrk 110:} Previous imaging studies revealed a tidal arm extending from the nucleus towards the west, suggesting a recent interaction or merger \citep[e.g.,][]{mackenty90, bischoff_kollatschny_99, veron_cetty_07}.
The \oiii~velocity map suggests large scale rotation, but a closer inspection of the \oiii~profile reveals multiple kinematic components due to the presence of double and triple-peaked profiles.
The gas dispersion map shows knots of higher line width towards the west and southwest, with the highest dispersion of $\sigma_{\text{gas}}\approx230$ km s$^{-1}$ corresponding to the west clump. 
It is in these high-dispersion clumps that we also find multi-peaked \oiii~line profiles.
We display the surface brightness within 60 km s$^{-1}$ bins in Figure \ref{fig:velslice}.
While there is indeed rotational motion depicted in the velocity channels, we also note the presence of several clumps located $\lesssim 10$ arcsec ($\lesssim 7.2$ kpc) from the nucleus.

The measured kinematic PA from the stellar velocity map suggests alignment with the larger-scale gas motion, but it should be noted that this galaxy has the largest uncertainty in PA$_\star$ across our sample.
Due to the projected proximity of a foreground star ($\sim 5$ arcsec away), the spectrum in many spaxels to the north-east of the nucleus is contaminated by starlight.
For these spaxels, we were unable to constrain the measurements of \mgib~lines from the host galaxy.

\textbf{Mrk 50:} With a masking threshold S/N of 3 for both the stellar and ionized gas kinematics, our maps only extend to $\sim 5$ arcsec (2.5 kpc) from the nucleus. 
The \oiii~emission is mainly elongated along the East-West direction, while the host galaxy emission is mainly oriented North-South.
We find no significant departure from rotation.

\textbf{Mrk 841:} The velocity map of the
\oiii~profile shows a distinct kinematic twist about the nucleus, extending out to $5$ arcsec (3.74 kpc) from the center. 
Within the extent of the kinematic twist, the PAs of the ionized gas and stellar velocity fields can differ by as much as $\Delta \rm{PA} \lesssim 90 \; \rm{deg}$.
The twist spatially coincides with an area of higher \oiii~velocity dispersion surrounding the nucleus and extending out to the same projected radius.
Along the major axis of the galaxy, we find high \oiii~line widths ($\sigma_{\text{gas}}\gtrsim200$ km s$^{-1}$), as well as profiles that have a slightly blueshifted wing.
For most spatial bins within 5 arcsec of the nucleus, we adopted a two-Gaussian decomposition for the \oiii~line profile, with a narrow core and a broad, asymmetric wing.
The velocity map for the narrow component (G1) exhibits a similar kinematic twist to the one displayed in Figure \ref{fig:full_b} (see Figure \ref{fig:sep_all}).
The velocity map for the asymmetric wing component (G2) shows a roughly East-West feature nearly 5 arcsec in radius, with with blueshifted gas located towards the east. 
The zero velocity line in the G2 velocity map is offset by $\sim 1.5$ arcsec towards the west of the nucleus.
Due to the differing kinematic structure of both \oiii~components, there are several regions where gas traced out by G1 and G2 are moving in opposite directions.

\textbf{Mrk 1392:} The \oiii~velocity map shows an s-shaped zero velocity line. This feature is still present in the maps for both components. 
The core component displays enhanced line widths in two regions located $\sim 5$ arcsec ($\sim 3.7$ kpc) away from the nucleus along the kinematic major axis. 
The spatial distribution of the non-rotational component displays two clumps extending towards the north east and south west, possibly tracing out the spiral arms seen in the imaging data.

\textbf{Zw 229-015:} The distribution of \oiii~is concentrated within 5 arcsec of the nucleus, with small packets of extended emission that partially trace out the spiral arms seen in imaging data. 
The ionized gas close to the nucleus is blueshifted, although the line profile contains narrow ($\sim 150$ km s$^{-1}$), symmetric line wings. 

The \sigstar~map displays an increase in velocity dispersion towards the nucleus, with maximum values of $\approx 90$ km s$^{-1}$ concentrated in a ring with a $\approx 2$ arcsec radius. 
The velocity dispersion drops in the immediate vicinity of the nucleus to $\approx 70$ km s$^{-1}$.
This behavior is similar to ``$\sigma$-drop'' galaxies, and has been seen in radial profiles of other AGN \citep[e.g.,][]{emsellem01, fraser-mckelvie_2024}.
This sudden decrease in \sigstar~may be caused by gas inflow onto a cold nuclear disk fueling star formation \citep{comeron08, fraser-mckelvie_2024}. 
Further analysis into the nature of Zw 229's apparent $\sigma$-drop requires high resolution observations focusing on the central kiloparsec, and is beyond the scope of this work.

\textbf{RX J2044.0$+$2833:} The \oiii~velocity map shows
a slight warp in the zero-velocity line when compared to the stellar velocity field,
while the line width map shows a streak of enhanced velocity dispersion roughly aligned with the photometric and kinematic axes. 

\textbf{Mrk 315:} Previous studies of Mrk 315 across different wavelengths point to a dwarf merger scenario \citep[e.g.,][]{mackenty94, nonino98, ciroi05}. 
We see a knot located 2 arcsec east of the nucleus with a stellar velocity of 700 km s$^{-1}$ relative to the systemic velocity, consistent with the findings for knot K in \cite{ciroi05}. 
The favored interpretation for knot K in \cite{ciroi05} is that it is leftover debris from a dwarf galaxy that had undergone a merger with Mrk 315.
The \oiii~profile shows two distinct kinematic components: one that follows disk rotation, and a second, highly redshifted ($v_{\text{gas}}\gtrsim400$ km s$^{-1}$) component that traces out a filamentary  structure running from northwest to southeast.
We detect the strongest emission on the east end of the filament, corresponding to region J1 seen in \oiii~maps in \cite{ciroi05}.
We display maps of the \oiii~surface brightness within 60 km s$^{-1}$ slices in Figure \ref{fig:velslice}, and note the excess flux in slices with velocities above $200$ km s$^{-1}$.

\begin{figure*}[htb!]
    \centering
    \includegraphics[width=\textwidth]{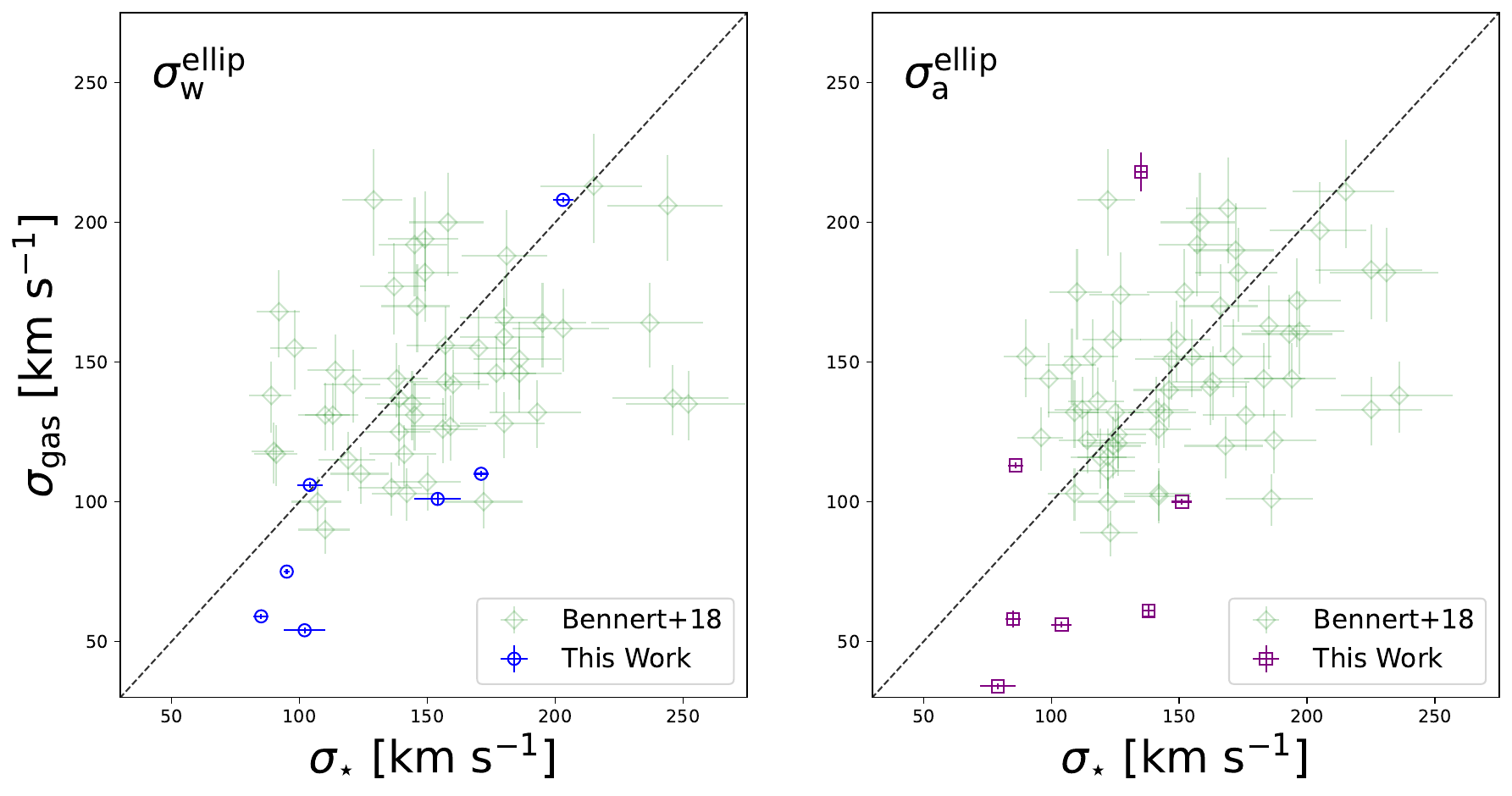}
    \caption{Comparison of \oiii~line width and \sigstar~for the flux-weighted dispersions (blue circles, left panel) and the fits to the aperture-summed spectra (purple squares, right panel) within an elliptical aperture.
    Data points based on measurements on local Seyfert 1 galaxies
    from \cite{bennert18}
    are shown
    as green
    diamonds.
    We adopt the velocity dispersion values of the two-Gaussian fit to \oiii~($\sigma_{\rm{[OIII],D}}$) for \siggas. 
    The dashed line is the 1:1 relation. For both methods of measuring the velocity dispersion, the majority of galaxies fall below the 1:1 line, suggesting that \siggas~underestimates \sigstar.}
    \label{fig:sig_comp}
\end{figure*}

\subsection{Aperture effects on velocity dispersion measurements}
The spatial information provided by our integral-field data enables us to test the assumption of circular symmetry used in previous studies with spatially resolved long-slit spectra. 
We obtained our aperture-wide values of $\sigma$ using an elliptical aperture defined by the best-fit parameters for the S\'ersic bulge.
In principle, measurements using the elliptical aperture should better represent the bulge kinematics, since the aperture is defined directly from the surface brightness photometry.
However, in the absence of integral-field data and precise measurements of the bulge morphology, the standard practice is to assume circular symmetry.
It is important to note that our ability to fully test differences between elliptical and circular apertures is limited by the seeing of our observations (typically 1\textendash2 arcsec) and the pixel scale of the KCWI data. 
Most objects in our sample have bulge effective radii comparable to or smaller than the seeing, meaning we cannot fully spatially resolve the bulge in our data.
In addition to the measurements listed in Table \ref{tbl:kin}, we also measured the flux-weighted ($\sigma_{\text{w}}$) and aperture-summed ($\sigma_{\text{a}}$) velocity dispersions for circular apertures of size $r_{\text{eff}}$.
We find that for the majority of our sample, the dispersions measured within a circular aperture generally agree within a few percent of the values measured within the elliptical aperture.
We do not expect any significant differences between our elliptical and circular aperture measurements since the bulges have nearly circular morphologies ($q \sim 0.8$), and given the spatial resolution of the data, elliptical apertures exclude only a few spaxels with lower flux that are located near the edge of the circular aperture.

We find significant differences between our elliptical aperture measurements and our pseudo long-slit measurements.
On average, the long-slit measurements deviate from the elliptical aperture measurements by 4 percent (standard deviation 4 percent, median 2 percent) for $\sigma_{\text{gas,w}}$, 13 percent (standard deviation 9 percent, median 11 percent) for $\sigma_{\star,\text{w}}$, 26 percent (standard deviation 27 percent, median 13 percent) for $\sigma_{\text{gas,a}}$, and 24 percent (standard deviation 29 percent, median 13 percent) for $\sigma_{\star,\text{a}}$.
While some galaxies show minimal differences of $\sim 1$ percent, others exhibit substantial deviations upwards of 40 percent.
Several factors may lead to the large differences seen in our dispersion measurements. 
Depending on how the size of the extraction window compares to that of the bulge, the long-slit extraction may exclude significant contributions from the bulge, or include non-bulge components.
Whether or not the light excluded (or included) by the long-slit extraction window affects measurements of the velocity dispersion further depends on the region's significance in the overall flux weighting, as well as how rapidly the spectrum profile changes.
The galaxy with the largest discrepancy in area between our long-slit extraction window and our elliptical aperture is Mrk 50, which has our largest angular $r_{\text{eff}}$ (4\farcs15). 
With our 0\farcs7 ``slit width'', the long-slit extraction window excludes the vast majority ($\sim$95 percent) of the bulge.
However, our results for Mrk 50 show that, depending on the measurement technique, the elliptical aperture and long-slit measurements only differ between 5 and 8 percent.
For the two targets where our long-slit extraction window is close to our elliptical apertures (Mrk 1392 and Zw 229-015), we still find differences of up to 30 percent.

Our tests show that for our sample, measurements from circular and elliptical apertures generally agree with each other within a few percent.
In contrast, our long-slit measurements can agree to within a few percent, or have significant differences exceeding $40$ percent relative to elliptical aperture measurements, which may depend on the target, dispersion measured, and measurement technique.

\subsection{Reliability of Stellar Kinematics from \mgib~and \ion{Fe}{1}} \label{subsec:mgfe}

We expect that AGN contamination will be the dominant source of uncertainty in our measurements of \sigstar.
To quantify the effects of AGN contamination, we investigate the fractional uncertainty in \sigstar~as a function of the AGN fraction of the integrated flux around $5100$ \AA. 

We find that for all fitted bins across the galaxy sample with AGN fractions $\geq 0.70$, the mean fractional error for measurements of \sigstar~is $8.3 \pm 4.9$ percent. 
However, since not every galaxy contains bins with high AGN fractions, this metric excludes four targets (Ark 120, MCG +04, Mrk 1392, and Mrk 315). 
If we increase the threhold for our analysis to bins with AGN fractions $\geq 0.80$, then the mean fractional error for measurements of \sigstar~is $10.9 \pm 6.5$ percent, which includes bins from only Zw 535, Mrk 110, and Mrk 841. 
Even within these three objects, some individual bins have uncertainties in \sigstar~of nearly $30$ percent.  

The extent to which AGN contamination affects our measurements of \sigstar~across the sample is object dependent. 
Within our ten galaxy sample, six contain bins with an AGN fraction $\geq 0.70$, while only three have bins with an AGN fraction $\geq 0.80$. We also emphasize that since these AGN fractions were derived in a spectral region around $5100$ \AA, these values likely underestimate the AGN fraction in the region around \mgib. 
Several lines of \feii~and [\feii] occur near $5170$ \AA, and thus mean that the AGN emission contributes more to the total flux near \mgib~than in the featureless region at $5100$ \AA.
Additionally, our analysis does not include the effects of emission lines such as [\ion{Fe}{6}] and [\ion{Fe}{7}] filling in and diluting the \mgib~absorption lines.

\begin{figure*}[htb!]
    \centering
    \includegraphics[width=\textwidth]{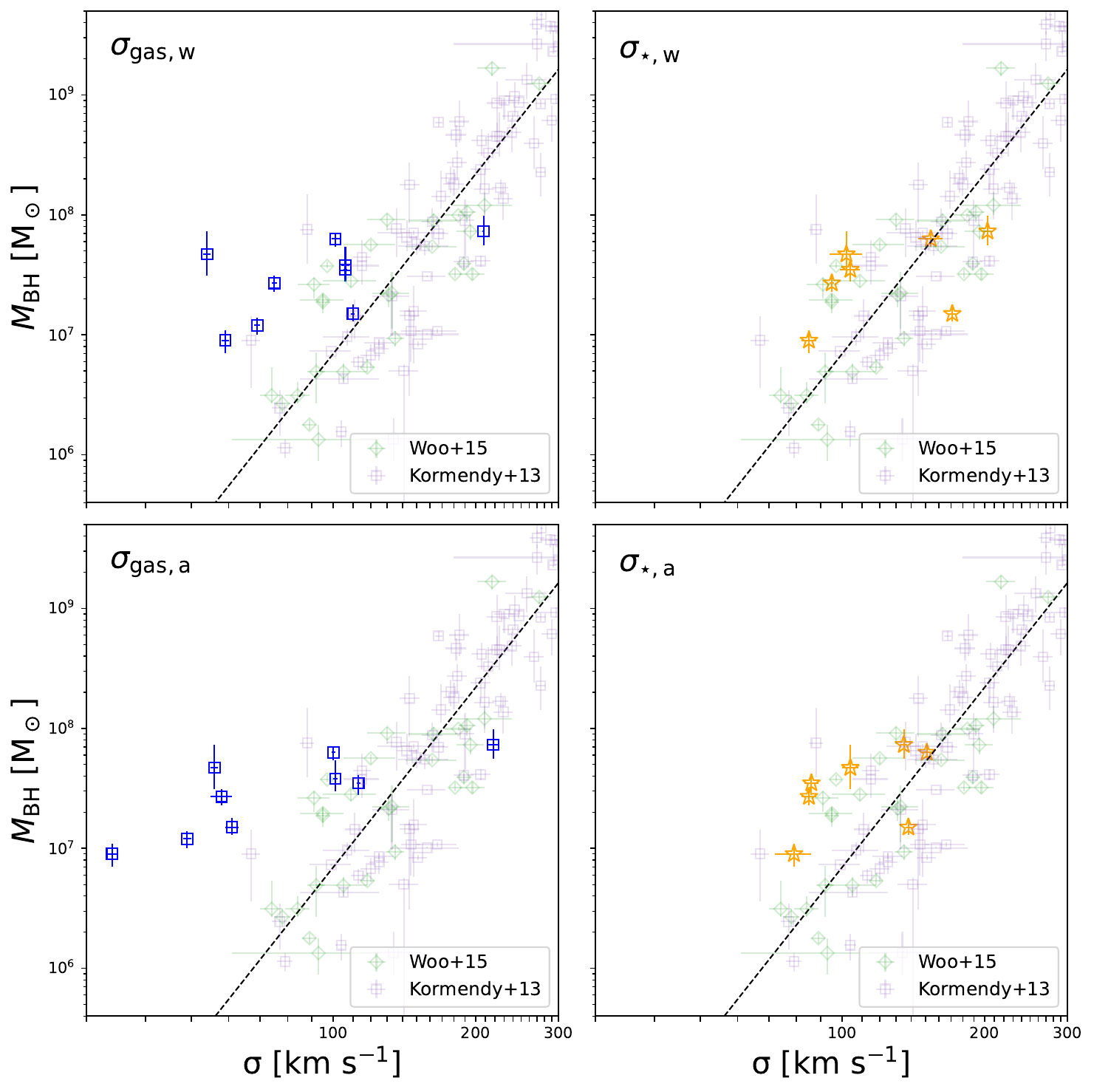}
    \caption{\mbh--$\sigma$ data points with \oiii~line widths (blue squares) and stellar velocity dispersions (orange stars) compared against data used in the analysis of \cite{woo15}. The dashed line is the best-fit \mbhsigstar~relation to a joint sample of quiescent galaxies from \cite{kormendy_ho_13} and reverberation-mapped AGN in \cite{woo15}.
    Plots in the top row correspond to the flux-weighted measurements $\sigma_{\text{w}}$, while those in the bottom row correspond to $\sigma_{\text{a}}$. The green diamonds are from the RM sample, while the gray squares correspond to the quiescent sample. Our $\sigma_{\text{gas}}$ tend to be lower than the range of the literature values for a fixed BH mass.}
    \label{fig:msig_gas}
\end{figure*}

\subsection{$\sigma_{gas}$ as a surrogate for \sigstar}
We compare the resulting dispersion measurements for the core component of \oiii~with the stellar velocity dispersions in Figure \ref{fig:sig_comp} (see also Table \ref{tbl:kin}).
Of the ten objects in our sample, we have measurements of $r_{\text{eff}}$ for nine targets obtained from surface brightness photometry on HST images (Mrk 315 does not have high resolution HST imaging data), and we obtain aperture-based \siggas~and \sigstar~measurements for seven targets.
For our flux-weighted measurements, we find a sample-wide mean dispersion ratio of $(\sigma_{\text{gas}} / \sigma_{\star})_{\text{w}} = 0.77$,
a standard deviation of $0.18$, a standard deviation of the mean of $0.07$, and a median dispersion ratio of $0.69$.
For the aperture-summed values, we measure a mean dispersion ratio of $(\sigma_{\text{gas}} / \sigma_{\star})_{\text{a}} = 0.81$, a standard deviation of $0.43$, 
a standard deviation of the mean of $0.18$, and a median dispersion ratio of $0.66$. 

Our results suggest that the \oiii~core line width generally underestimates \sigstar, with our flux-weighted measurements differing from $(\sigma_{\text{gas}} / \sigma_{\star})_{\text{w}} = 1.0$ by more than $3\sigma$, although our aperture-summed values only differ by $\sim 1\sigma$.
We performed additional measurements of $\sigma_{\text{gas}}$ and $\sigma_{\star}$ using circular apertures equal to the size of the galaxy effective radius $r_{\text{gal}}$ as described in \cite{winkel25}.
While these apertures yielded different values of \siggas~and \sigstar, they do not affect the overall result that within our sample, $\sigma_{\text{gas}}$ is generally smaller than $\sigma_{\star}$.
This is in contrast to other studies that have carried out this comparison, where the dispersions ratios are more consistent with 1.0 \citep[e.g.,][]{greene_ho_05, bennert18}. 
Galaxies with our most discrepant data points (e.g., MCG $+$04-22-042, Mrk 50, Mrk 841, Mrk 1392, Zw 229-015) contained \oiii~profiles that required very thin cores to fit the profile.
With the exception of Mrk 50, these galaxies displayed non-rotational kinematics in their \oiii~velocity maps, and even in their velocity maps for the isolated G1 component. The discrepancy between \siggas~and \sigstar~has been shown to be affected by AGN activity or the presence of outflows, suggesting that these may have to be taken into account when evaluating the viability of using the \oiii~line width as a substitute for the stellar velocity dispersion \citep[e.g.,][]{sexton21, le23}. 

\cite{greene_ho_06} showed that narrow and broad AGN emission can bias \sigstar~measurements obtained from \mgib.
In cases where AGN emission contaminates \mgib, \cite{greene_ho_06} advise using the \ion{Fe}{1} region ranging from 5250 \AA ~to 5820 \AA.
However, given the 5600 \AA ~cutoff of the KCWI blue channel and the redshift range of our targets, this translates to rest-frame red limits from $5340$ \AA \textendash $5470$ \AA.
Due to our instrumental limitations, our wavelength coverage does not extend redward enough to include $5600$ \AA \textendash $5800$ \AA, where \feii~emission is weaker.
The addition of KCWI red channel observations covering \ion{Fe}{1}, and more importantly the \ion{Ca}{2} triplet will enable more thorough testing of \sigstar~measurements, as well as the use of \siggas~as a proxy for \sigstar.

Our analysis suggests that for our sample of galaxies, the width of the \oiii~line core does not serve as an accurate tracer for the stellar velocity dispersion. We consider the possibility that the interpretation of the core-wing decomposition of the \oiii~line profile may not be sufficient for some objects. 
As previously mentioned, for most objects in this sample, we restricted our \oiii~fits to use at most two Gaussians, and interpreted the narrower, higher amplitude core as tracing the dynamically relaxed motion in the gravitational potential of the host galaxy. 
This method, while previously used in studies involving nearby ($z\leq0.1$) AGN hosts \citep[e.g.,][]{greene_ho_05, bennert18, sexton21}, yields values of \siggas~that underestimate \sigstar~when applied to our sample.
One possible explanation for the discrepancy between \sigstar~and \siggas~ is the interplay between structure and orientation.
If the ionized gas has settled into a flattened disk, then when viewed face-on, one can expect to measure lower line-of-sight velocity dispersions compared to motions of stars in a spheroidal bulge. 
While this may explain why most of our objects lie below the 1:1 line in Figure \ref{fig:sig_comp}, it is also important to note that the gas may adopt various spatial distributions and orientations relative to the host galaxy. This also means that comparisons between \siggas~and \sigstar~ are highly object-dependent, and thus \siggas~is not reliable as a precise estimator of \sigstar~for individual objects.
However, when applied to statistical samples of AGN, \siggas~measurements may still provide useful constraints on scaling relations that traditionally utilize \sigstar.

\subsubsection{Use of \oiii~line width in \mbhsigstar~studies}

In this work, we investigate the use of the \oiii~kinematics as a tracer of the stellar kinematics in \mbhsigstar~studies.
We plot the \oiii~core line widths over \mbh--\sigstar~points found in the literature in Figure \ref{fig:msig_gas}. 
We note that visually, for a fixed value of \mbh, our measurements of \sigstar~generally lie within the range of stellar velocity dispersion values seen in the samples of quiescent galaxies from \cite{kormendy_ho_13} and reverberation-mapped AGN from \cite{woo15}.
In contrast, our measurements of \siggas~tend to be biased towards lower values, and appear to lie outside the horizontal scatter of the \sigstar~measurements from the literature.
While the \mbhsigstar~relation is discussed in detail in \cite{winkel25}, we restrict our discussion in this paper to the \oiii~kinematics. 
A more rigorous statistical analysis of \siggas~as a surrogate for \sigstar~in the context of \mbhsigstar~will be explored with a larger sample of reverberation-mapped galaxies in a future paper. 

\subsubsection{Possible use of FWHM/2.355}
We revisit the original comparison made in \cite{nelson_whittle_96} and \cite{nelson00}, which used the FWHM of the \oiii~line profile instead of the velocity dispersion of the core component.
Empirically, using FWHM/2.355 of the full line profile provided the best tracer for \sigstar~across the entire sample.
We find a mean ratio of $1.04$ with a sample standard deviation of $0.38$, 
a standard deviation of the mean of $0.16$, and a median of $0.93$. 
Our result suggests that FWHM/2.355 of the full line profile may be a more reliable tracer of \sigstar~compared to \siggas, but the large scatter indicates that this must be done across an ensemble of galaxies, rather than on an individual basis. 
A major difference between the use of FWHM/2.355 from the full \oiii~profile and the width of the line core as a tracer for $\sigma_{\star}$ is in attributing the contribution of the gravitational or virialized component to the full profile.
Using FWHM/2.355 operates on the assumption that within the bulge, the gas is primarily driven by the bulge gravitational potential. 
In contrast, using the width of the line core requires one to assign physical meanings to separate Gaussians in the spectral decomposition, which can lead to complications if the \oiii~profile is fitted with three or more Gaussians, or in cases where the rotational component could have a non-Gaussian shape. 
Testing the viability of FWHM/2.355 as another surrogate for \sigstar~will require observations across a larger sample of galaxies with diverse \oiii~profiles.

\begin{figure*}[htb!]
    \includegraphics[width=\textwidth]{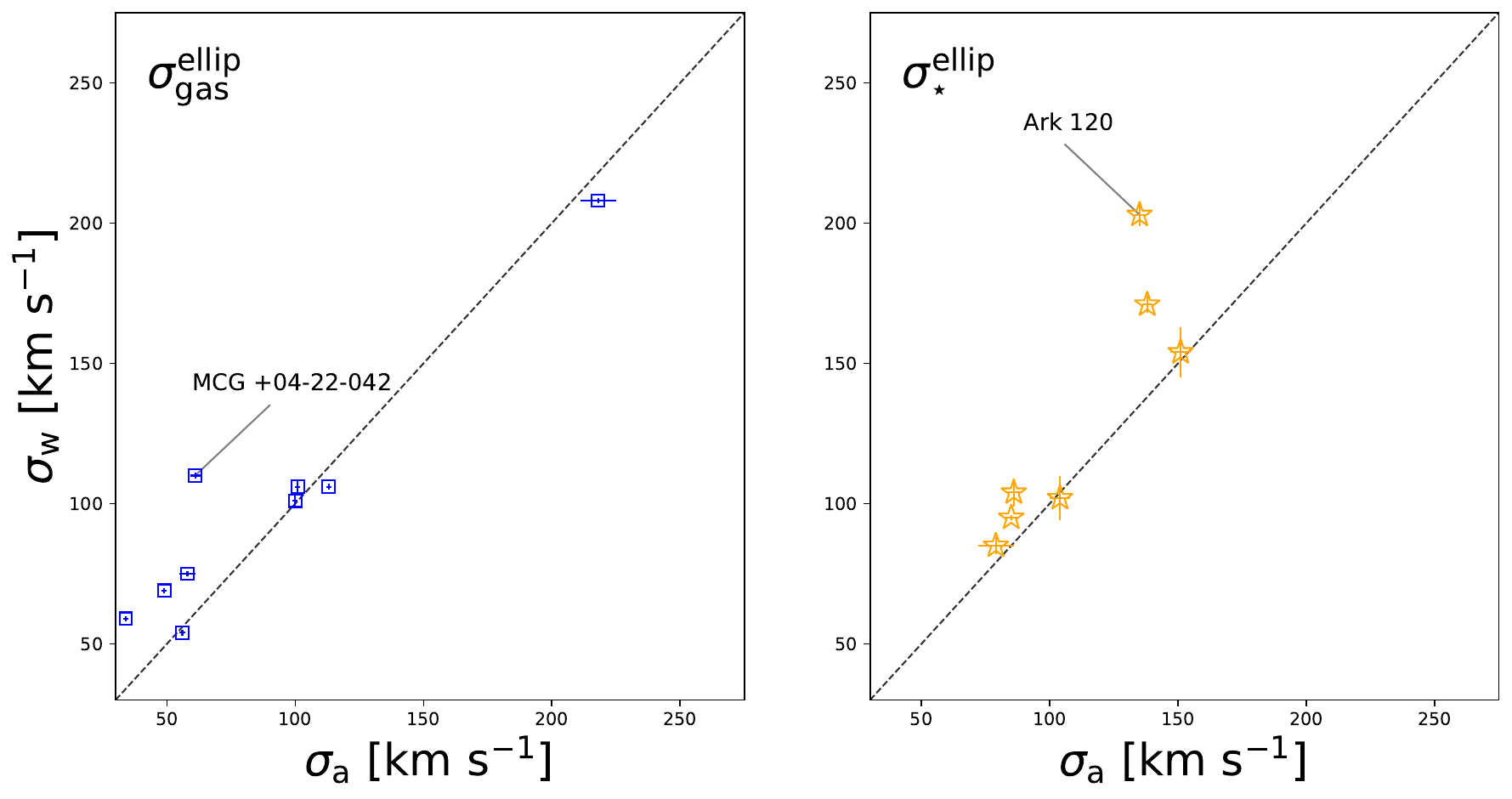}
    \caption{Comparison between the flux-weighted and aperture-summed measurements of velocity dispersion for the gas (left panel) and the stars (right panel). The dashed line represents equality between $\sigma_{\text{w}}$ and $\sigma_{\text{a}}$. 
    We denote targets where $\sigma$ measurements between both methods are notable outliers from the 1:1 relation.}
    \label{fig:sig_w_a_comparison}
\end{figure*}

\subsection{Aperture-summed vs Flux-weighted $\sigma$ measurements} \label{sub:avsw}

We directly compare measurements between the aperture-summed ($\sigma_{\text{a}}$) and flux-weighted ($\sigma_{\text{w}}$) dispersion measurements in Figure \ref{fig:sig_w_a_comparison}.
Between the different measurements of the \oiii~line width, most objects have $\sigma_{\text{gas,a}} \leq \sigma_{\text{gas,w}}$, with MCG$+$04-22-042 displaying the largest discrepancy.
For this object, $\sigma_{\text{gas}} \sim 100$ km s$^{-1}$ for individual bins within the elliptical aperture, while the best-fit line width to the aperture-summed spectrum is closer to $60$ km s$^{-1}$.
Across our sample, we measure a mean ratio $\sigma_{\text{gas,w}} / \sigma_{\text{gas,a}} = 1.24$,
a standard deviation of $0.32$,
a standard deviation of the mean of $0.11$, and a median ratio of $1.05$. 
Evaluating this ratio for \sigstar~yields a mean ratio of $1.16$, 
a standard deviation of $0.16$, 
a standard deviation of the mean of $0.07$, and a median ratio of $1.12$.
The one outlier in this comparison is Ark 120, where $\sigma_{\star,\text{w}} = 1.50 \sigma_{\star,\text{a}}$, which can be attributed to the high AGN contamination within our elliptical aperture.

A slight systematic difference between $\sigma_{\text{a}}$ and $\sigma_{\text{w}}$ is expected. 
In calculating the aperture flux-weighted dispersion $\sigma_{\text{w}}$, the velocity dispersion in the $i$-th bin $\sigma_{i}$ is calculated relative to the line centroid $\lambda_i$, and $v_i$ is calculated relative to the rest wavelength $\lambda_0$.
For $\sigma_{\text{a}}$, both $v_i$ and $\sigma_i$ are calculated relative to the centroid wavelength $\lambda_\text{c}$ of the aperture-summed profile. 
In the absence of a rotational velocity gradient (that is, if $\lambda_i=\lambda_c=\lambda_0$ for all bins within the aperture), then both methods would yield the same measurement of the velocity dispersion.
In effect, this means that the contribution of each individual bin to $\sigma_{\text{a},v}^2$ is dependent on the aperture-wide velocity field, while contributions to $\sigma_{\text{w},v}^2$ are independent of it.

While the above demonstrates that $\sigma_{\text{w}}$ and $\sigma_{\text{a}}$ are fundamentally different, it is not clear in which situations $\sigma_{\text{w}} > \sigma_{\text{a}}$ and vice-versa. Studies with integral-field data have shown that there is a tendency for $\sigma_{\star,\text{w}} < \sigma_{\star, \text{a}}$ \citep{kormendy_ho_13, batiste17}.
Studies using long-slit data have also shown that the discrepancy between $\sigma_\text{w}$ and $\sigma_\text{a}$ is dependent on inclination, velocity amplitude, as well as the aperture size \citep{son20}.
\cite{bennert15} notes that for their sample of local Seyfert 1s, the aperture-summed values of \sigstar~agree within a few percent relative to their spatially resolved values.
If we attribute this difference purely due to the different contributions of $v_i$ and $\sigma_i$ between the definitions of $\sigma_\text{w}$ and $\sigma_\text{a}$, then we would expect to see similar differences for \siggas, assuming a similar (rotational) velocity field for the gas within the bulge.
For our sample, differences between $\sigma_{\star,\text{a}}$ and $\sigma_{\star,\text{w}}$ are 13 percent on average, while some individual galaxies have differences of over 30 percent. 
We obtained similar values for $\sigma_\text{gas}$, with differences of 14 percent on average, with some individual galaxies having differences exceeding 40 percent.

A more thorough comparison between $\sigma_\text{w}$ and $\sigma_\text{a}$ will be the topic of a future study with a larger sample of galaxies.

\begin{deluxetable*}{l c c c c c c c c}[htb]
\tablecaption{NLR Size and $L_{\text{[O III]}}$ Measurements} 
\tablecolumns{9}
\tablewidth{3pt}
\tablehead{
\colhead{Object} & 
\colhead{$R_{\text{NLR}}^{\text{ratio}}$} &
\colhead{$R_{\text{NLR}}^{\text{SB}}$} &
\colhead{$R_{\text{NLR}}^{\text{ratio,half}}$} &
\colhead{$R_{\text{NLR}}^{\text{SB,half}}$} &
\colhead{$L_{\text{[O III]}}^{\text{ratio}}$} &
\colhead{$L_{\text{[O III]}}^{\text{SB}}$} &
\colhead{$R_{\text{NLR}}^{\text{Gauss}}$} &
\colhead{$L_{\text{[O III]}}^{\text{Gauss}}$}
\\
\colhead{} &
\colhead{(kpc)} &
\colhead{(kpc)} &
\colhead{(kpc)} &
\colhead{(kpc)} &
\colhead{($10^{41}$ erg s$^{-1}$)} &
\colhead{($10^{41}$ erg s$^{-1}$)} &
\colhead{(kpc)} &
\colhead{($10^{41}$ erg s$^{-1}$)}
}
\startdata
Zw 535-012        & $2.98\pm1.18$ & $2.36\pm0.31$ & $0.75\pm0.02$ & $0.74\pm0.01$ & $2.20\pm0.05$ & $2.16\pm0.03$ & $0.28\pm0.02$ & $1.73\pm0.08$\\
Ark 120           & $2.29\pm0.33$ & $3.34\pm0.33$ & $1.27\pm0.10$ & $1.48\pm0.04$ & $2.52\pm0.29$ & $3.18\pm0.14$ & $0.40\pm0.02$ & $2.78\pm0.11$ \\
MCG $+$04-22-042  & $>5.60$ & $6.76\pm1.31$ & $>1.96$ & $2.04\pm0.10$ & $>4.85$ & $5.07\pm0.25$ & $0.17\pm0.04$ & $3.60\pm0.12$ \\
Mrk 110           & $>8.40$ & $7.75\pm0.79$ & $>2.23$ & $2.18\pm0.06$ & $>8.89$ & $8.67\pm0.29$ & $0.30\pm0.02$ & $5.96\pm0.16$ \\
Mrk 50            & $1.51\pm0.17$ & $1.44\pm0.20$ & $1.00\pm0.08$ & $0.92\pm0.16$ & $0.08\pm0.01$ & $0.07\pm0.02$ & $\cdots$ & $\cdots$  \\
Mrk 841           & $>6.40$ & $3.96\pm0.20$ & $>1.68$ & $1.59\pm0.02$ & $>10.0$ & $9.37\pm0.15$ & $0.25\pm0.02$ & $7.39\pm0.23$ \\
Mrk 1392          & $5.43\pm0.66$ & $4.69\pm0.94$ & $1.59\pm0.03$ & $1.55\pm0.08$ & $7.73\pm0.15$ & $7.55\pm0.40$ & $0.21\pm0.03$ & $6.63\pm0.22$ \\
Zw 229-015        & $2.58\pm0.59$ & $1.50\pm0.15$ & $1.05\pm0.09$ & $0.78\pm0.06$ & $0.82\pm0.07$ & $0.59\pm0.06$ & $0.09\pm0.04$ & $0.71\pm0.06$\\
RX J2044.0$+$2833 & $3.21\pm0.95$ & $3.36\pm0.53$ & $0.65\pm0.02$ & $0.65\pm0.01$ & $4.15\pm0.11$ & $4.17\pm0.07$ & $0.15\pm0.03$ & $2.87\pm0.15$ \\
Mrk 315           & $0.93\pm0.12$ & $7.77\pm3.63$ & $0.58\pm0.06$ & $1.87\pm1.65$ & $1.44\pm0.21$ & $4.94\pm1.47$ & $0.20\pm0.02$& $2.52\pm0.14$ \\
\enddata \label{tbl:nlr_size}
\tablecomments{ $R_{\rm{NLR}}^{\text{ratio}}$ and $R_{\rm{NLR}}^{\text{SB}}$ correspond to the NLR extents derived from the [\ion{O}{3}]/H$\beta$ flux ratio and the \oiii~surface brightness cutoff, respectively. 
The corresponding values of $L_{\text{[O III]}}$ refer to the total \oiii~luminosity located within a circular aperture of radius $R_{\text{NLR}}$. 
$R_{\text{NLR}}^{\text{ratio,half}}$ and $R_{\text{NLR}}^{\text{SB,half}}$ are the half-light radii containing half of the \oiii~luminosities specified in the fourth and fifth columns. 
We report lower limits MCG$+$04-22-042, Mrk 110, and Mrk 841 for the ratio-based measurements since [\ion{O}{3}]/H$\beta$ stayed consistently above the threshold.
The final two columns correspond to $R_{\text{NLR}}$ and $L_{\text{[O III]}}$ as measured by a Gaussian fit to the \oiii~surface brightness profile. $L_{\text{[O III]}}$ is derived from the total flux contained within the Gaussian.
Since our initial measurement of $R_{\text{NLR}}^{\text{Gauss}}$ was less than the seeing for Mrk 50, we do not report a measurement for this object.
}
\end{deluxetable*}

\begin{figure*}[ht!]
    \includegraphics[width=\textwidth]{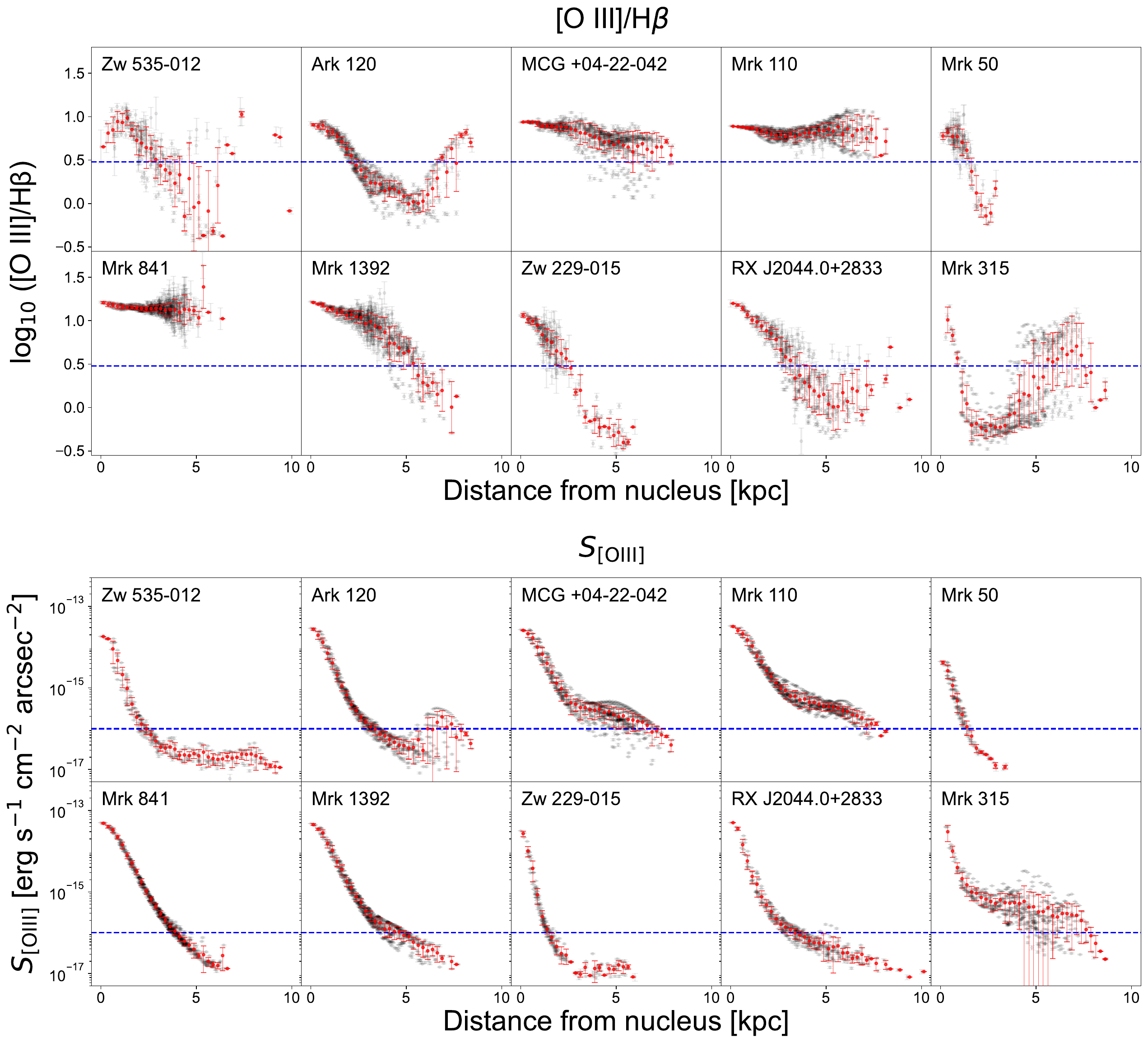}
    \caption{\oiii/H$\beta$ flux ratio (top) and \oiii~surface brightness (bottom) as a function of projected distance from the nucleus. 
    The black points represent the flux ratio or surface brightness in a spatial bin whose centroid is located some distance from the nucleus.
    Each red point denotes the mean flux ratio for all black points in a given 0.25 kpc bin from the nucleus.
    The red error bars are the standard deviations calculated from all the black points in a given distance bin. The blue dashed line represents the threshold of \oiii/H$\beta$ $=3$ or $S_{\text{[O III]}} = 10^{-16}$ erg s$^{-1}$ cm$^{-2}$ arcsec$^{-2}$.}
    \label{fig:nlr_profile}
\end{figure*}

\begin{figure*}[ht!]
    \centering
    \includegraphics[width=0.75\textwidth]{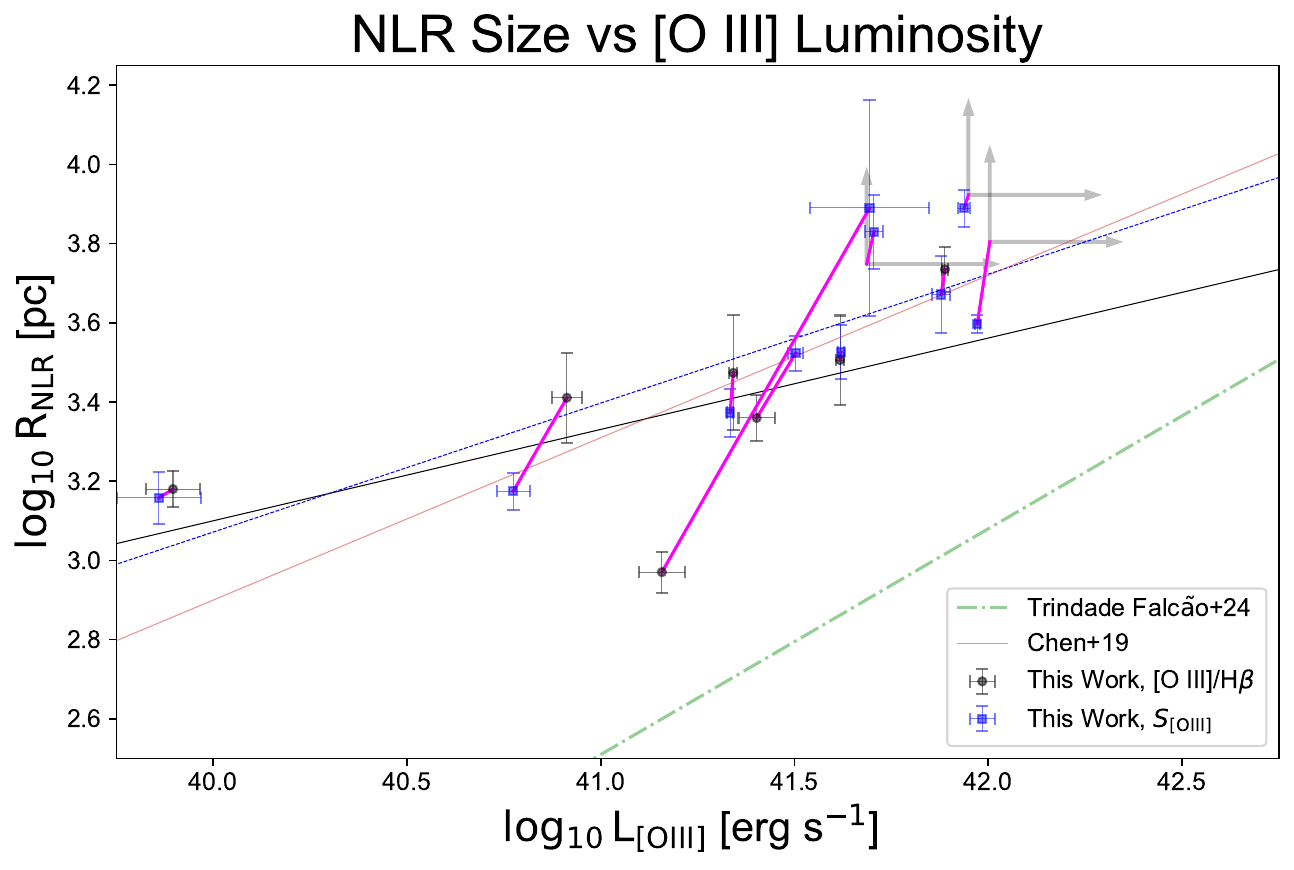}
    \caption{The $R_{\text{NLR}}-L_{\text{[O III]}}$ relation for our sample. 
    Our data points based on the [\ion{O}{3}]/H$\beta$ ratio are plotted as black circles and gray arrows (for lower limits), with the best-fit relation as the solid black line. 
    Our points based on the \oiii~surface brightness cutoff are plotted as blue squares, with the best-fit relation as the dotted blue line. 
    The pink lines connect measurements associated with the same galaxy.
    For comparison, we plot the \citet{falcao2024} relation based on HST imaging data as the green dot-dashed line, and the \citet{chen19} relation from SDSS/MaNGA integral-field data as the solid red line. }
    \label{fig:nlr_size_lum}
\end{figure*}

\section{The extent of the AGN Narrow-Line Region} \label{sec:nlr}

An additional application of our data is measuring the extent of the AGN narrow-line region. 
Previous studies have measured the size of the NLR from imaging \citep[e.g.,][]{bennert02, schmitt03, falcao2024}, long-slit \citep[e.g.,][]{fraquelli03, bennert06, greene11}, and integral-field \citep[e.g.,][]{husemann13, husemann14, liu14, chen19} observations, and have also revealed a correlation between the extent of the NLR and the corresponding \oiii~luminosity $L_{\text{[O III]}}$.
However, there is no standard definition for the size of the NLR across the literature, with definitions based on surface brightness cutoffs \citep[e.g.,][]{bennert02, schmitt03, falcao2024} and emission line ratios \citep[e.g.,][]{bennert06, chen19}, or luminosity-weighted radii \citep[e.g.,][]{husemann13, husemann14}.
This is accompanied by the use of both ``narrow-line region'' (NLR) and ``extended narrow-line region'' (ENLR) in the literature to refer to the region where the gas is primarily ionized by the AGN \citep[e.g.,][]{bennert06, husemann13}. 
While there is no specific physical criterion for distinguishing the NLR from the ENLR, past work has based the distinction on whether the emission extends past 1 kpc \citep[e.g.,][]{husemann13, chen19}.
This in turn leads to different interpretations of the measured NLR sizes, as well as the derived $R_{\text{NLR}}$--$L_{\text{[O III]}}$ scaling relations.
With our high quality integral-field data from KCWI, we can directly compare how different $R_{\text{NLR}}$ definitions applied to the same dataset affect the measured NLR sizes.

\subsection{Measurements of the NLR size}

We base our first set of $R_{\text{NLR}}$ measurements on the [\ion{O}{3}]/H$\beta$ flux ratio as measured from the full line profiles. 
We define $R_{\text{NLR}}$ for each galaxy as the distance from the nucleus at which the \oiii$/$H$\beta$ ratio drops below $3$ (equivalently, $\log_{10}$([\ion{O}{3}]$/$H$\beta$) $<0.48$). 
This \oiii$/$H$\beta$ threshold was chosen since regions with [\ion{O}{3}]/H$\beta \geq 3 $ are likely dominated by AGN photoionization \citep[e.g.,][]{ho97, husemann13}, and a drop below this value may indicate a transition to ionization dominated by \ion{H}{2} regions \citep[e.g.,][]{bennert06, greene11}.
We first plot the \oiii$/$H$\beta$ ratio as a function of the projected distance from the nucleus for each spatial bin.
We then group the data points into 0.25 kpc bins, and calculate the mean flux ratio and standard deviation over the binned points (Figure \ref{fig:nlr_profile}, top panel).
From the radial profiles, we perform a linear interpolation between profile data points, and find the distance at which \oiii$/$H$\beta$ falls below $3$. 
We obtain $L_{\text{[O III]}}$ by summing up the flux for all bins located within a circular aperture of radius $R_{\text{NLR}}$.
Repeating this process for the upper and lower limits of the line ratios, we derive (if possible) uncertainties for $R_{\text{NLR}}$ and $L_{\text{[O III]}}$. 
For targets where we could not obtain all three measurements for $R_{\text{NLR}}$ and $L_{\text{[O III]}}$ (MCG$+$04-22-042, Mrk 110, and Mrk 841), we report their values as lower limits. 

For comparison with imaging-based studies that only use a surface brightness cutoff, we performed a second set of measurements, this time defining $R_{\text{NLR}}$ as the distance the \oiii~surface brightness drops below $10^{-16}$ erg s$^{-1}$ cm$^{-2}$ arcsec$^{-2}$.
We adopt this threshold for the purpose of comparison with the work of \cite{chen19}.
Similarly to our [\ion{O}{3}]/H$\beta$-based measurements, we create a surface brightness radial profile (Figure \ref{fig:nlr_profile}, bottom panel). 

In addition to the two previous definitions of the narrow-line region size, we also measure the half-light radii of the high [\ion{O}{3}]/H$\beta$ region ([\ion{O}{3}]/H$\beta$ $>3$) and the \oiii-bright region ($S_{\text{[O III]}} > 1 \times 10^{-16}$ erg s$^{-1}$ cm$^{-2}$ arcsec$^{-2}$). 
That is, we measure the radius that encompasses half of the \oiii~luminosity obtained from our previous $R_{\text{NLR}}$ measurements.
This measurement provides a more compact, characteristic size of the NLR containing the higher surface brightness \oiii~emission. 

We also measure the size of the NLR based on the innermost 1\textendash2 kpc of the \oiii~surface brightness radial profiles in Figure \ref{fig:nlr_profile}. 
We modeled the surface brightness profile as a single Gaussian plus a constant, and calculated the size of the NLR from the width of the Gaussian.
To account for the effects of atmospheric seeing, we subtracted the seeing divided by 2.355 in quadrature from the Gaussian $\sigma$, and report the result as $R_{\text{NLR}}^{\text{Gauss}}$.
This particular definition of $R_{\text{NLR}}$ encompasses the brightest and most compact regions of the \oiii~emission, and is particularly relevant for studies where the narrow-line emission is not spatially resolved (e.g., lensed quasars at high redshift; \citealt{nierenberg20}).
We report our measurements in Table \ref{tbl:nlr_size}.

\subsection{Dependence on measurement technique}

We obtain $R_{\text{NLR}}$ measurements using both the isophotal radius and [\ion{O}{3}]/H$\beta$ ratio for seven of our galaxies. Of these, four (Zw 535-012, Mrk 50, Mrk 1392, RX J2044.0+2833) show agreement within uncertainties between NLR sizes measured with both techniques, while some show discrepancies of nearly an order of magnitude.
Ark 120 yields a larger NLR size measurement if the \oiii~surface brightness cutoff is used, but this can be explained if there is bright \oiii~emission even if the emission line ratio is below our chosen threshold.
Thus, for measurements based purely on a surface brightness cutoff, it is possible for one to overestimate the the size of the NLR, because one could be including regions that are not dominated by AGN photoionization.
We obtain a smaller NLR size for Zw 229-015 if we use the surface brightness technique instead of the [\ion{O}{3}]/H$\beta$ ratio.
This can be explained by our chosen surface brightness threshold of $10^{-16}$ erg s$^{-1}$ cm$^{-2}$ arcsec$^{-2}$.
With this chosen threshold, our measured NLR size excludes some of the fainter regions that are still likely photoionized by the AGN.
The largest discrepancy between our NLR measurements across different techniques is seen in Mrk 315.
The [\ion{O}{3}]/H$\beta$ ratio quickly drops below the threshold at $\sim 1$ kpc and thus yields our smallest measurement. 
However, Mrk 315 also yields our largest NLR measurement based on the \oiii~surface brightness cutoff due to the highly extended emission from the merger remnant. 

Our measured half-light radii reveal that the strongest \oiii~emission is concentrated within $\sim 2$ kpc of the AGN, with several objects (depending on measurement technique) having sub-kpc half-light radii. 
For both the line ratio-based and surface brightness-based measurements, the half-light radii are on average 40 percent of their corresponding values of $R_{\text{NLR}}$. 
RX J2044.0$+$2833, has the smallest half-light radii relative to its $R_{\text{NLR}}$, with most of its higher surface brightness emission concentrated in a region that is only $20$ percent of its NLR size. 
In contrast, we measure half-light radii for Mrk 50 that are $\gtrsim60$ percent of their corresponding $R_{\text{NLR}}$ values, indicating a more extended distribution of the high surface brightness \oiii~emission. 

The NLR sizes based on the Gaussian fits to the \oiii~surface brightness profile correspond to the most compact regions we measure. 
We measure typical values of $R_{\text{NLR}}^{\text{Gauss}} \gtrsim 100$ pc, indicating that across our sample, the brightest \oiii~emission is concentrated within a few hundred pc of the nucleus.

We were unable to measure $R_{\text{NLR}}$ for three targets (MCG $+$04-22-042, Mrk 110, Mrk 841) based on the [\ion{O}{3}]/H$\beta$ ratio, since the radial profiles consistently stayed above the threshold at all distances. 
The spatial distribution of these three objects extends to the edge of the KCWI FoV, so it is likely that the size of the NLR extends well beyond the KCWI FoV.
Additionally, with our flux ratio-based definition of $R_{\text{NLR}}$, targets with highly extended emission may still have their NLR sizes underestimated. 
Ark 120 and Mrk 315 display distinct increases in their [\ion{O}{3}]/H$\beta$ radial profile at larger distances, reaching the end of the KCWI FoV, but our measurements only correspond to the first dip below [\ion{O}{3}]/H$\beta$ $= 3$.
A possible modification to our definition is to instead find the largest distance at which the radial profile transitions from above the threshold to below it, but this will require observations that cover a much larger FoV.

\subsection{Comparison to other studies}

For comparison with other studies studying the extent of the NLR, we used \texttt{CapFit}\footnote{\url{https://www-astro.physics.ox.ac.uk/~cappellari/software/}} \citep{cappellari23} to obtain the linear relations
\begin{multline*}
\log_{10}\left( \frac{R_{ \text{NLR} }}{\text{pc}}\right) = (0.23\pm0.14) \log_{10} \left( \frac{L_{\text{[O III]}}}{\text{erg sec}^{-1}}\right) \\
- (6.10\pm5.79)
\end{multline*}
for our measurements based on the [\ion{O}{3}]/H$\beta$ line ratio, and 
\begin{multline*}
\log_{10}\left( \frac{R_{ \text{NLR} }}{\text{pc}}\right) = (0.33\pm0.08) \log_{10} \left( \frac{L_{\text{[O III]}}}{\text{erg sec}^{-1}}\right) \\
- (9.95\pm3.16)
\end{multline*}
for our measurements based on the \oiii~surface brightness cutoff. 
Due to our limited sample size and sparse sampling in luminosity, we will restrict our discussion to the NLR sizes we measure for a given luminosity.
We display our results in Figure \ref{fig:nlr_size_lum}. 
Immediately, we note a large discrepancy between our measured NLR sizes and the relation based on HST imaging data of far-infrared-selected Seyferts and radio-quiet type 1 quasars \citep{schmitt03,falcao2024}. 
For our luminosity range, we measure NLR sizes 1-2 orders of magnitude larger than expected from their relation.
This discrepancy is likely explained by the depth of the \citet{schmitt03} data used to expand the luminosity range of the \citet{falcao2024} paper to lower values. 
The \citet{schmitt03} data reaches depths of several $10^{-15}$ erg s$^{-1}$ cm$^{-2}$, which is an order of magnitude brighter than our surface brightness cutoff, and about two to three orders of magnitude brighter than the depth of our observations. 
If we apply this threshold for our data, then our resulting NLR sizes would be typically only a few kpc, excluding much of the extended emission.

We compare our measurements to the \citet{chen19} study, which measured $R_{\text{NLR}}$ for $\sim150$ BPT-selected AGN hosts from the SDSS/MaNGA survey. 
While a direct comparison is not possible due to our different definitions of $R_{\text{NLR}}$, our NLR sizes still fall close to their derived relation.
\citet{chen19} use a combination of both BPT diagnostics and a surface brightness cutoff, while our definitions consider the line ratio and the surface brightness cutoff separately.
Applying both techniques simultaneously to our data would effectively mean adopting the smaller of our two $R_{\text{NLR}}$ measurements for a given galaxy.
Additionally, we calculate $L_{\text{[O III]}}$ by integrating over all spaxels within a circular aperture of $R_{\text{NLR}}$, while \citet{chen19} only integrate over spaxels dominated by AGN photoionization.
Despite the differences in our methods for determining $R_{\text{NLR}}$ and $L_{\text{[O III]}}$, our results still agree with the \citet{chen19} relation to within 0.5 dex.

Our integral-field data from KCWI reveal that AGN photoionization can affect the host galaxy out to distances of $\sim 10$ kpc. 
With our deeper observations, we measure NLR extents that are larger than the sizes found by past HST imaging data \citep[e.g.,][]{schmitt03}, but are consistent with measurements from other integral-field observations \citep[e.g.,][]{chen19}.
In addition, the nature of integral-field data enables direct comparisons between different NLR measurement techniques (e.g., emission line diagnostics, surface brightness thresholds) for a homogeneous data set, without the limitations of long-slit (e.g., dependence on slit orientation) or imaging (e.g., lack of spectral information) data.

\section{Summary}\label{sec:summary}
In this paper, we presented integral-field spectroscopy for our pilot sample of ten local Seyfert 1 galaxies obtained from KCWI at the Keck II Telescope.
From the high-quality integral-field data, we measured and extracted spatially resolved kinematics of the stars and gas.
We investigated the alignment between the global kinematic PAs of the stars and gas, and explored the viability of using the \oiii~line width as a surrogate for \sigstar.
In addition, we investigated how different measurement techniques affect the measured extent of the AGN narrow-line region.
We summarize our results as follows:
\begin{enumerate}
    \item The large-scale \oiii~kinematics are dominated by rotation at galaxy-wide spatial scales. We do, however, find that most objects show signatures of non-rotational motion near the nucleus (e.g., kinematic twist or s-shaped warps in the velocity field).
    
    \item The global kinematic PA for the \oiii~velocity field is aligned with that of the stellar velocity ($\Delta \text{PA} \leq 30$ deg). 
    The larger scale motion is dominated by rotation, and non-rotational motion near the nucleus has minimal effect on measurements of the global kinematic PA.
    
    \item Comparisons of the line width of the narrow \oiii~core component to the stellar velocity dispersion show that our data points lie mostly below the 1:1 relation, with average dispersion ratios of $0.77$ and $0.81$ and standard deviations of the mean of $0.07$ and $0.18$ for the flux-weighted and aperture-summed measurements, respectively. This suggests that the width of the \oiii~line core is not an accurate tracer of the stellar velocity dispersion. We find that FWHM/2.355 as measured from the full \oiii~profile yields a sample-wide average ratio of $1.04$, a standard deviation of $0.38$, 
    a standard deviation of the mean of $0.16$.
    Our measurements suggest that FWHM/2.355 of the \oiii~profile may serve as a more viable proxy for \sigstar.     

    \item We measured AGN NLR sizes ranging from $\sim 0.1$ to $\sim 10$ kpc across the four definitions of $R_{\text{NLR}}$ we adopted.
    For a given value of $L_{\text{[O III]}}$, our measurements of $R_{\text{NLR}}$ based on either the surface brightness cutoff or the [\ion{O}{3}]/H$\beta$ line ratio are an order of magnitude larger than the sizes based on past HST imaging data, due to the higher sensitivity of our deep integral-field observations.
    
\end{enumerate}

The integral-field observations from KCWI have enabled us to map the stellar and ionized gas kinematics for a sample of ten galaxies with precisely determined BH masses, and measure the extent of their narrow-line regions. 
Our analysis has highlighted the importance of deep integral-field data in measuring the size of the NLR, but also in exploring and comparing different definitions used in determining $R_{\text{NLR}}$.
We intend to apply our analyses in this paper to a larger sample of 29 local 
($ 0.01 < z < 0.09$) galaxies with precisely determined BH masses spanning several orders of magnitude, enabling a more systematic study of the gas and stars in AGN hosts. \linebreak

RR would like to thank Justin Kader and Marina Bianchin for helpful and engaging discussions regarding spectral fitting, as well as the interpretation of the kinematic maps.

RR and VU acknowledge funding support from STScI grant \# HST-AR-17063.005-A and NSF Astronomy and Astrophysics Grant (AAG) \# AST-2408820.
VU further acknowledges support from NASA Astrophysics Data Analysis Program (ADAP) grant \# 80NSSC23K0750, ADSPS grant \# 80NSSC25K7477, and STScI grants \# HST-GO-17285.001-A and \# JWST-GO-01717.001-A, which were provided by NASA through a grant from the Space Telescope Science Institute, which is operated by the Association of Universities for Research in Astronomy, Inc., under NASA contract NAS 5-03127.
VNB gratefully acknowledges support from the ESO Scientific Visitor Program.
VNB acknowledges assistance from the National Science Foundation (NSF) through grant AST-1909297. Support for Program number HST-GO 17103 (PI Bennert) and HST-AR 17063 (PI Bennert) was provided through a grant from the STScI under NASA contract NAS5-26555.

Some of the data presented in this paper were obtained from the Mikulski Archive for Space Telescopes (MAST) at the Space Telescope Science Institute. The specific observations analyzed can be accessed via \dataset[https://doi.org/10.17909/63n4-8k32]{https://doi.org/10.17909/63n4-8k32}. STScI is operated by the Association of Universities for Research in Astronomy, Inc., under NASA contract NAS5–26555. Support to MAST for these data is provided by the NASA Office of Space Science via grant NAG5–7584 and by other grants and contracts.

This research has made use of the NASA/IPAC Extragalactic Database (NED),
which is operated by the Jet Propulsion Laboratory, California Institute of Technology,
under contract with the National Aeronautics and Space Administration.

The data presented herein were obtained at the W. M. Keck Observatory, which is operated as a scientific partnership among the California Institute of Technology, the University of California and the National Aeronautics and Space Administration. The Observatory was made possible by the generous financial support of the W. M. Keck Foundation.

This research has made use of the Keck Observatory Archive (KOA), which is operated by the W. M. Keck Observatory and the NASA Exoplanet Science Institute (NExScI), under contract with the National Aeronautics and Space Administration.

Funding for the Sloan Digital Sky Survey V has been provided by the Alfred P. Sloan Foundation, the Heising-Simons Foundation, the National Science Foundation, and the Participating Institutions. 
SDSS acknowledges support and resources from the Center for High-Performance Computing at the University of Utah. The SDSS web site is \url{www.sdss.org}.

SDSS is managed by the Astrophysical Research Consortium for the Participating Institutions of the SDSS Collaboration, including the Carnegie Institution for Science, Chilean National Time Allocation Committee (CNTAC) ratified researchers, the Gotham Participation Group, Harvard University, Heidelberg University, The Johns Hopkins University, L’Ecole polytechnique f\'{e}d\'{e}rale de Lausanne (EPFL), Leibniz-Institut f\"{u}r Astrophysik Potsdam (AIP), Max-Planck-Institut f\"{u}r Astronomie (MPIA Heidelberg), Max-Planck-Institut f\"{u}r Extraterrestrische Physik (MPE), Nanjing University, National Astronomical Observatories of China (NAOC), New Mexico State University, The Ohio State University, Pennsylvania State University, Smithsonian Astrophysical Observatory, Space Telescope Science Institute (STScI), the Stellar Astrophysics Participation Group, Universidad Nacional Aut\'{o}noma de M\'{e}xico, University of Arizona, University of Colorado Boulder, University of Illinois at Urbana-Champaign, University of Toronto, University of Utah, University of Virginia, Yale University, and Yunnan University.

%\clearpage

%\vspace{5mm}
\facilities{Keck: II (KCWI)}


\software{
astropy (The Astropy Collaboration 2013, 2018),
BADASS (Sexton et al. 2019, 2021),
lenstronomy (Birrer \& Amara 2018, Birrer et al. 2021),
Matplotlib (Hunter 2007),
NumPy (Harris et al. 2020),
PAFit (Krajnovi\`{c} et al. 2006),
PyPipe3D (Lacerda et al. 2022),
SciPy (Virtanen et al. 2020),
VorBin (Cappellari \& Copin 2003)}



\bibliography{kcwi_lamp_paper}{}
\bibliographystyle{aasjournal}



\end{document}